\documentclass[
 reprint,
 amsmath,amssymb,
 aps,
floatfix
]{revtex4-1}
\usepackage[T2A]{fontenc}

\usepackage{graphicx}
\usepackage{dcolumn}
\usepackage{bm}

\begin{document}

\preprint{APS/123-QED}

\title{ Calculation of thermal conductivity coefficients of electrons in  magnetized dense matter}

\author{G. S. Bisnovatyi-Kogan}%
 \email{gkogan@iki.rssi.ru}
\affiliation{Space Research Institute, Rus. Acad. Sci., Profsoyuznaya str. 84/32,  Moscow 117997, Russia.}
\affiliation{National Research Nuclear University MEPhI (Moscow Engineering Physics Institute),
Kashirskoye Shosse 31, Moscow, 115409, Russia.}
\author{M. V. Glushikhina}
\email{m.glushikhina@iki.rssi.ru}
\affiliation{Space Research Institute, Rus. Acad. Sci., Profsoyuznaya str. 84/32,  Moscow 117997, Russia.}



\date{ \today}

\begin{abstract}

The solution of Boltzmann equation for plasma in magnetic field, with arbitrarily
degenerate electrons and non-degenerate nuclei, is obtained by Chapman-Enskog method.
Functions, generalizing Sonin polynomials are used for obtaining an approximate solution.
Fully ionized plasma is considered. The tensor of the heat
conductivity  coefficients in non-quantized magnetic field is
calculated.  For  non-degenerate and strongly degenerate plasma the
asymptotic analytic formulas are obtained, which are compared with
results of previous authors.
The Lorentz approximation, with neglecting of electron-electron encounters, is asymptotically
exact for strongly degenerate plasma.

 We obtain, for the first time, in three polynomial approximation,
with account of electron-electron collisions,
analytical expressions for the heat conductivity tensor for
non-degenerate electrons, in presence of a magnetic field. Account of the third polynomial improved substantially the precision  of results.
 In two polynomial approximation our solution coincides with the published results.

For strongly degenerate electrons we obtain, for the first time, an asymptotically exact analytical solution  for the heat conductivity tensor in presence of a magnetic field. This solution has considerably more complicated dependence on the magnetic field than those in previous publications, and gives  several times smaller relative value of a thermal conductivity across the magnetic field at $\omega\tau\gtrsim 0.8$.

\end{abstract}

\maketitle


\section{\label{sec:level1}Introduction}

Observations of thermal emission from neutron stars (NS) provides
information about the magnetic
field strength and configuration,
temperature, chemical composition of the outer regions,
and about the properties of matter at higher
densities, deeper inside the star see ~\cite{pagegep06},\cite{ponswallat02}.
To derive this information, we need to calculate the structure and
evolution of the star, and compare theoretical models with
observational data.
X-ray observations of thermal emission show periodic variabilities in single neutron stars ~\cite{zav2007},
indicating to the anisotropic
temperature distribution. It is produced  at the low  and intermediate
density regions, such as the solid crust, where a complicated
magnetic field geometry could cause a coupled magneto-thermal
evolution. In some extreme cases, with a very high magnetic field, this anisotropy may even be
present in the poorly known interior, where  neutrino processes are
responsible for the energy removal \cite{km10}.

The spectrum of these NSs in broad range from optics
to X-ray band, cannot be reproduced by a spectrum of the surface
with a unique temperature.
Fitting of the spectrum of the X-ray source     $   RXJ 1856.53754$ in
this broad region is explained  ~\cite{ponswallat02}   by a small hot
emitting area  $10 - 20$ $ {\rm km}^2$, and an extended cooler
component. Another piece of evidence that strongly supports the
nonuniform temperature distribution are pulsations in the X-ray
signal of some objects of amplitudes  $5 - 30\,\%$, some of which
have irregular light curves that point towards a non-dipolar
temperature distribution \cite{agponsmir08}.

Heat transfer in the envelopes of NS  plays crucial role
in many aspects of evolution  of these stars. Thermal conductivity
is the basic quantity needed for calculating the  relationship
between the internal temperature of a neutron star and its effective
surface temperature. This relationship affects thermal evolution of
the neutron star and its radiation spectra. To calculate thermal
conductivity we should know the transport properties of a dense matter
where electrons are degenerate, and form a nearly ideal Fermi-gas \cite{yaklevshib1999}.
The ions are usually treated as non-degenerate. They may be in a gaseous state, may form a  Coulomb liquid or
a Coulomb crystal \cite{salpeter1961}. Under such conditions, electrons are the most
important  heat carriers, and the thermal conductivity is determined
by electron motion. The magnetic field limits the motion of
electrons in  directions perpendicular to the field lines and, since
they are the main carriers of the heat transport, the thermal
conductivity in these directions is suppressed, while
remaining  unaffected along the field lines.
The conductivity of  electrons in NS and white dwarfs
in presence of a magnetic
field was studied in  ~\cite{fl76},~\cite{yau80}.  The ratio
between thermal conductivity along and across magnetic field lines
considered in \cite{fl76} was taken as

\begin{equation}
 \label{firstone}
 \frac{\lambda_{\perp}}{\lambda_{\parallel}} = \frac{1}{1+ (\omega \tau)^{2}}.
 \end{equation}
and was used also in \cite{yau80}. Here $\omega$ is electron cyclotron frequency,
$\tau$ is the time between collisions.
The influence
of the magnetic field on the electron heat conductivity in the form
 (\ref{firstone}) was used in  subsequent papers, see
~\cite{itohhayashi93},
~\cite{ygklp}.
Here we find an analytic solution for the heat conductivity tensor of strongly degenerate electrons
in a magnetic field, in the Lorentz approximation, which is asymptotically exact in this case, showing
a more complicated dependence on the magnetic field strength than (\ref{firstone}).

Classical methods of kinetic gas theory were developed by Maxwell, Boltzmann,
Gilbert, Enskog and Chapman.  These methods are presented in the
monograph of Chapman and Cowling ~\cite{chap90}.  They  are based
on the solution of the Boltzmann equation by method of successive
approximations. As a zeroth
approximation, the thermodynamic equilibrium distribution function is taken,
which is a Maxwell distribution for a non-degenerate gas, and a Fermi-Dirac
distribution in cases when degeneracy is important.
The equilibrium distribution function is not an  exact solution of Boltzmann equation in presence of non-uniformity.
 Following \cite{chap90} we look for a
solution of the Boltzmann equation in the first approximation by
Sonyne  (Laguerre) polynomial expansion, valid for a non-degenerate gas.
To take into account a degeneracy, we use a set of orthogonal functions, as a generalization
of Sonyne polynomials, suggested in \cite{uu}, \cite{uel}, \cite{tom}, see also
\cite{bkrom82}.
Only two first
 terms of this expansion are taken often for  calculations of the
heat conductivity coefficients.  It was shown in \cite{imsh} that this approximation
gives substantial errors for the coefficient of a heat conductivity, which
become much smaller when 3 polynomial expansion is used. Here we calculate
the heat conductivity tensor in a magnetic field using  3 terms of the expansion.
We show the improvement of the precision, when mowing from 2 to 3 polynomials, on the example
of the non-magnetized Lorentz gas, where an exact solution is
known.

The first application of the Boltzmann equation  to the gas of
charged particles was made by Chapman
~\cite{chap90}.   Due to the divergence of the collision integral
at large impact parameters for particles with the Coulomb interaction,
the upper limit of integration  over the impact parameter was taken
at the length of the average distance between  particles. Thus the
coefficients of viscosity, heat conductivity and diffusion of gases
composed of charged particles were obtained. Divergence of the
collision integral for Coulomb interaction at large impact parameter
shows that the scattering with large impact parameters and small
change of momentum in one collision play more important role than
collisions with big change of the momentum. Landau used this fact to
simplify the  Boltzmann collision integral \cite{landau37}. He expanded distribution
function after collision in small variations of momentum and left
first two terms. The integral obtained in  this way is called Landau
collision integral. Another derivation of the Landau collision
integral was done by Chandrasekhar \cite{chandra}, who used the
analog with the Brownian motion, described by the Fokker-Planck
equation. The identity between Landau and Fokker-Planck collision
integrals was shown in \cite{rmj}, see also \cite{trub}.

The kinetic coefficients in the non-degenerate plasma, with and without magnetic field had been
calculated in \cite{lands51}, \cite{marsh90}, \cite{bk64},  \cite{bkl88} using Chapmen-Enskog expansion method .
Braginskii ~\cite{brag57} calculated the kinetic coefficients
for a  non-degenerate  plasma in magnetic field consisting of electrons and one
sort of positively charged ions, using kinetic equations normalized to  average velocities, different for the ions and electrons.  Landau
collision integral was used, and two polynomials were taken into account in the expansion.
The same approach, was used in ~\cite{bs93}, where
calculations  of kinetic coefficients for a fully ionized plasma of
a complex composition have been performed.
Coefficients of the heat conductivity tensor in a degenerate stellar
cores  were calculated  in  Lorentz approximation for a hydrogen plasma in \cite{wyller}.
A nonrelativistic calculation, based on on the quantum Lenard-Balescu transport equation for the thermal and electrical
conductivity of plasma of highly degenerate, weakly coupled electrons and nondegenerate, weakly coupled ions was performed in \cite{lampe}.

Shear viscosities of non-relativistic, relativistic and ultra-relativistic hard sphere gas  were calculated
by Chapman-Enskog method in \cite{WiranataPrakash2012},\cite{wir1} 

The present work is devoted to the  solution of the  Boltzmann
equation by Chapman-Enskog method for electrons in an
arbitrary degenerate plasma. We find a tensor of the heat conductivity
using the expansion in two and three polynomials, and, on the example of the
Lorentz gas, we show that the
method has a good convergence to the exact solution.

 We obtain, for the first time, in three polynomial approximation,
with account of electron-electron collisions,
analytical expressions for the heat conductivity tensor for
non-degenerate electrons, in presence of a magnetic field. Account of the third polynomial improved substantially the precision of results.
 In two polynomial approximation our solution coincides with the published results.

For strongly degenerate electrons we obtain, for the first time, an asymptotically exact analytical solution  for the heat conductivity tensor in presence of a magnetic field. This solution has considerably more complicated dependence on the magnetic field than those in previous publications, and gives  several times smaller relative value of a thermal conductivity across the magnetic field at
 $\omega\tau\gtrsim 0.8$.

\section{Boltzmann equations and transfer equations}

We use a Boltzmann
equation for  electrons, in a magnetic field,
with  an allowance of arbitrary degeneracy,
and assuming them as non-relativistic
We consider  the electron gas in a crystal lattice of
heavy nuclei, and  take into account the interaction of the
electrons with a nondegenerate nuclei and with one another. The
nuclear component of the matter in the crust is  in a crystal
state, and therefore the isotropic part of the distribution function
$f_{N0}$ may differ from the Maxwellian distribution. If the mass of
the nucleus $m_{N}$ is much greater, than the electron mass $m_{e}$,
then to the terms $\sim m_{e}/m_{N}$ the details of the distribution
function $f_{N0}$ are unimportant, and the calculations can be made
for arbitrary  $f_{N0}$.

Boltzmann equation, which describes the time variation of the
electron  distribution function $f$ in presence of the electric and
magnetic  fields is written as \cite{bk64}, \cite{bkl88}

\begin{equation}
 \label{Boltz_gen}
\frac{\partial f}{\partial t} + c_{i}\frac{\partial f}{\partial r_{i}}
- \frac{e}{m_e}(E_{i}+ \frac{1}{c}\varepsilon_{ikl} c_{k} B_{l})
\frac{\partial f}{\partial c_{i}} + J = 0.
\end{equation}
Here $(-e) , m_e$ are the charge (negative) and the mass of the electron, $E_{i},
B_{i}$ are the strength of the electric field, and magnetic
induction, $ J$ is a collision integral, $\varepsilon_{ikl}$ is the
totally antisymmetric Levi-Civita tensor, $c$ is the speed of the
light.

\begin{eqnarray}
 \label{stoss}
\begin{aligned}
J = J_{ee} + J_{eN}  = R \int [f^{'}f_{1}^{'}(1-f)(1-f_{1})- \\
- ff_{1}(1-f^{'})(1-f^{'}_{1})]\times g_{ee}b\,db\, d\varepsilon dc_{1i}+\\
+ \int[f^{'}f^{'}_{N}(1-f) - ff_{N}(1-f^{'}) ] \times g_{eN}b\,db\, d\varepsilon dc_{Ni}.
\end{aligned}
\end{eqnarray}
Here, the impact parameter $b$, and $\varepsilon$ are geometrical
parameters of particle collisions  with relative velocities

$$g_{ee},\quad g_{eN};\qquad R=\frac{2m_e^3}{h^3}.$$
 The integration in electron part of the collision integral in (\ref{stoss}) is performed over the phase space
of the incoming particles ($dc_{1i}$), and the physical space of their arrival ($b\,db\,d\varepsilon$) \cite{chap90}.
 The velocity functions after collision are
marked with touches.

The Boltzmann equation for electrons with a binary collision integral (\ref{stoss}) may be applied in conditions, when
the electron gas may be considered as almost ideal, i.e. the kinetic energy of the electrons is much larger than
the energy of electrostatic interactions. It is valid for plasma at sufficiently small density. In the neutron stars and white dwarfs we
have an opposite conditions of plasma at very large density, when it is important to take into account the electrons degeneracy. It is
known from the statistical physics, that a gas of strongly degenerate electrons becomes ideal, because large Fermi energy substitutes here the
thermal energy \cite{dau90}. Therefore the calculations in this paper are applied to the low density, and high density plasma with degenerate electrons.
Detailed discussion of the applicability of a binary collision integral (\ref{stoss}), and its modifications for high density non-degenerate gases
may be found in \cite{chap90}.

Lets introduce  the thermal velocity of  electrons, $v_{i} =
c_{i}-c_{0i}$, where $c_{0i}$ is the mass-average velocity.
 So we can write the Boltzmann equation with
respect to the thermal velocity in the form \cite{bkl88}

\begin{eqnarray}
 \label{Boltz}
\begin{aligned}
\frac{df}{dt} + v_{i}\frac{\partial f}{\partial
r_{i}} - \left[ \frac{e}{m_e}(E_{i}+ \frac{1}{c}\varepsilon_{ikl} v_{k}
B_{l}) + \frac{dc_{0i}}{dt}\right]  \frac{\partial f}{\partial
v_{i}} \\
-  \frac{ e}{m_e c} \varepsilon _{ikl} v_{k} B_{l}
\frac{\partial f}{\partial v_{i}} -\frac{\partial
f}{\partial v_{i}}v_{k} \frac{\partial c_{0i}}{\partial r_{k}}+ J =0,
\end{aligned}
\end{eqnarray}
where
$$\frac{d}{dt}=\frac{\partial}{\partial t}+c_{0i}\frac{\partial}{\partial r_i}.
$$
The transfer equations for the electron concentration, total
momentum,  and electron energy, in the two-component mixture of electrons and
nuclei, can be obtained in a usual manner  from the
Boltzmann equation in a quasi-neutral plasma ~\cite{chap90,
marsh90, bk64, bkl88} as

\begin{equation}
 \label{electron number}
\frac{dn_{e}}{dt}+n_{e}\frac{\partial c_{0i}}{\partial r_i}
+\frac{\partial }{\partial r_{i}}(n_{e}\langle   v_{i} \rangle)=0,
\end{equation}

\begin{equation}
 \label{transport2}
\rho
\frac{dc_{0i}}{\mathit{dt}}=\frac 1 c\varepsilon _{\mathit{ikl}}j_kB_l
-\frac{\partial \Pi_{ik}}{\partial r_k},
\end{equation}

\begin{eqnarray}
 \label{energy}
\begin{aligned}
\frac{3}{2}{kn_{e}}\frac{{dT}}{{dt}} - \frac{3}{2}kT\frac{\partial}
{\partial r_{i}}(n_{e}\langle v_{i}\rangle)+\frac{\partial q_{ei}}
{\partial r_{i}}+ \Pi_{ik}^e\frac{\partial c_{0i}}{\partial r_k}
=\\=j_{i}(E_{i}+\frac{1}{c}\varepsilon_{ikl}c_{0k}B_{l}) - \rho_{e}
\langle v_{i}\rangle\frac{dc_{0i}}{dt},
\end{aligned}
\end{eqnarray}
where:

\begin{eqnarray}
 \label{pressure}
\begin{aligned}
\Pi_{ik} &   =\sum\limits_\alpha{ n_\alpha m_\alpha\langle v^\alpha_{i} v^\alpha_{k}\rangle},  & \Pi_{ik}^e &
= n_{\alpha} m_{\alpha}\langle v_{i} v_{k}\rangle,
\end{aligned}
\end{eqnarray}

\begin{eqnarray}
\begin{aligned}
\langle v_{ \alpha i} \rangle & = \frac{R}{n_{\alpha}} \int f v_{\alpha i} dc_{\alpha i},
 & n_e& = R\int f dc_{ei},
\end{aligned}
\end{eqnarray}

\begin{eqnarray}
\begin{aligned}
c_{0i} & = \frac{1}{\rho} \sum\limits_\alpha{\rho_{\alpha}\langle c_{ai} \rangle},
&j_{i} & =- n_{e}e\left\langle v_{i} \right\rangle,
 \end{aligned}
\end{eqnarray}

\begin{equation}
\label{qi}
 q_{ \alpha i}
= \frac{1}{2}n_{\alpha} m_\alpha \langle v_{\alpha}^{2}v_{\alpha i} \rangle.
\end{equation}
Here summation is taken over the electrons and nuclei, $\Pi_{ik}^e=P_e\delta_{ik}$,
$P_e=\frac{1}{3}n_e m_e \langle v^{2}\rangle$, when we neglect the electron viscosity,
$P_{e}$  is the electron pressure, $\langle v_{i} \rangle$  is in average
electron velocity in the comoving system, $q_{i}$ is the electron heat flux, and $j_{i}$
is the electron electric current. Here and in the subsequent consideration we
identify the mass average velocity with the average velocity of ions
$c_{0i}= \langle c_Ni\rangle$, and we consider the electric
current and heat flux produced only by electrons. In the quasi neutral plasma the electron
concentration $n_{e}$ is uniquely connected with the density $\rho$, defined by $\langle A,Z \rangle$ nuclei, $m_{N} = Am_{p}$

\begin{eqnarray}
\begin{aligned}
\rho & = m_{N}n_{N},
 & n_e& = \frac{Z\rho}{m_{N}}.
\end{aligned}
\end{eqnarray}

\section{Derivation of general equations for the first approximation function}

The Boltzmann equation can be solved by Chapmen-Enskog method of
successive approximation ~\cite{chap90}.
This method is used here for conditions, when distribution functions are close to their values in thermodynamic equilibrium,
and deviations are considered in a linear approximation. Equation for second order deviation from the equilibrium distribution function had been
derived in \cite{burnett} for a simple gas, see also \cite{chap90}. The complexity of this equation, and rather narrow region where second order corrections
could be important, strongly restricted the application of this approach.

The zeroth approximation to
the electron distribution function is a Fermi-Dirac distribution,
which is found by equating to zero of the collision integral $J_{ee}$ from  ~(\ref{stoss})

\begin{equation}
 \label{dist_func}
f_0  =[1+\exp \left(\frac{m_e v^2-2\mu}{2\mathit{kT}}\right)]^{-1}, \quad R\int{f_0dv_i}=n_e.
\end{equation}
Here, $\mu$ is a chemical potential of electrons, $k$ is Boltzmann's
constant,  $T$ is the temperature.
The nuclear distribution function
in the zeroth approximation  $f_{N0}$ is assumed to be isotropic
with respect to the  velocities and to depend on the local
thermodynamic parameters; otherwise it can be arbitrary with the
normalization:

\begin{equation}
 \label{nuclear_norm}
n_N=\int f_{\mathit{N0}}\mathit{dc}_{Ni},
\end{equation}
where $n_N$ is the concentration of nuclei, $n_e=Z\,n_N$, $Z$ is the
charge of the nucleus. Using ~(\ref{dist_func}) in~(\ref{electron
number})-(\ref{qi}),  we obtain the zeroth approximation for
the transfer equations. In this approximation $\langle v_{i} \rangle
=0,\,\,\, q_{i} = 0,\,\,\, \Pi_{ik} = (P_{e}+ P_{N})\delta_{ik}$

\begin{equation}
 \label{el_num-den}
\begin{aligned}
n_e & =2  \left( \frac{2kTm_{e}}{h^{2}}\right) ^{3/2} G_{3/2}(x_0),\\
 P_e &= 2\mathit{kT} \left( \frac{2kTm_{e}}{h^{2}}\right)^{3/2}G_{5/2}(x_0),
\end{aligned}
\end{equation}

\begin{eqnarray}
\label{fermiintegral}
G_n(x_0)=\frac
{1}{\Gamma(n)}\int_0^\infty{ \frac{x^{n-1}dx}{1+exp(x-x_{0})} },\,\,\,
x_0=\frac{\mu }{\mathit{kT}},
\end{eqnarray}
where $G_n(x_0)$ are Fermi integrals.  In what follows, instead of
$G_{n}(x_{0})$ we will write $G_{n}$ cause the argument is the same.
In the first approximation, we seek for the function $f$ in the
form:

\begin{equation}
 \label{chi}
f=f_0[1+\chi(1-f_0)].
\end{equation}
The distribution function $f_{N0}$ is  assumed to satisfy the relation:

\begin{equation}
 \label{normalization}
\frac{1}{n_{N}} \int v_{Ni} v_{Nk}f_{N0}dc_{Ni} = \delta_{ik} \frac{kT}{m_{N}}.
\end{equation}
The function $\chi$ admits representation of the solution in the form:

\begin{equation}
\label{chi_a}
\begin{aligned}
\chi & = -A_{i}\frac{\partial \ln T}{\partial r_{i}}-n_{e}D_{i}d_{i}\frac{G_{5/2}}{G_{3/2}}, \\
\end{aligned}
\end{equation}

\begin{equation}
\label{chi_b}
d_{i} = \frac{\rho_{N}}{\rho} \frac{ \partial \ln P_{e}}{\partial r_{i}}
- \frac{\rho_{e}}{P_{e}}\frac{1}{\rho}\frac{\partial P_{N}}{\partial r_{i}}
+\frac{e}{kT}(E_i+\frac{1}{c} \varepsilon _{ikl} c_{0k} B_{l}).
\end{equation}
The plasma is supposed to be quasineutral with a zero charge density.
The functions $A_{i}$ and $D_{i}$ determine the  heat
transfer  and diffusion. Substituting ~(\ref{chi_a})  in the
equation for $\chi$  we  obtain equations for $A_{i}$,
$D_{i}$ \cite{chap90}. It was shown in
~\cite{bk64},\cite{bkl88}, that in presence of a magnetic field
$B_{i}$, the polar vector $A_i$ (and similarly $D_i$) may be searched for in the form:

\begin{equation}
\begin{aligned}
A_{i}  & = A^{(1)}v_{i}+A^{(2)} \varepsilon_{ijk}v_{j}B_{k}+A^{(3)}B_{i}(v_{j}B_{j}),  \\
 \end{aligned}
\end{equation}
 Introducing a function:

\begin{equation}
\label{xicomp}
\xi  = A^{(1)}+iBA^{(2)},
\end{equation}
and dimensionless velocity: $u_{i} =
\sqrt{\frac{m_{e}}{2kT}}v_{i}$, we obtain the system for $\xi$ as

\begin{widetext}
\begin{equation}
\label{eq_sys}
\begin{aligned}
 f_{0}(1-f_{0})(u^{2} - \frac{5G_{5/2}}{2G_{3/2}})u_{i}  =
\frac{i}{3}  \frac{em_{e} B}{ \rho k T c}u_{i}f_{0}(1-f_{0})\left[\int\xi_{N} f_{N0}v_{N}^{2}dc_{Ni}
-   R\int \xi f_{0}(1-f_{0})v_{e}^{2}dc_{i}\right] \\
-iBf_{0}(1-f_{0})\frac{e \xi}{m_{e} c}u_{i}+ I_{ee}(\xi u_{i})+I_{eN}(\xi_{Ni} u_{Ni}),
\end{aligned}
 \end{equation}
\end{widetext}
where

\begin{eqnarray}
I_{ee}(\xi u_{i}) = R\int f_{0}f_{01}(1-f^{'}_{0})(1-f_{01}^{'})\times (\xi u_{i}\\
\nonumber +\xi_{1}u_{1i}-\xi^{'} u_{i}^{'}-\xi_{1}^{'}u^{'}_{1i})g_{ee}bdbd\varepsilon dc_{1i},
\end{eqnarray}

\begin{eqnarray}
I_{eN}(\xi u_{Ni}) = \int f_{0}f_{N0}(1-f^{'}_{0})(\xi u_{i} \\ \nonumber -\xi^{'} u_{i}^{'})g_{eN}bdbd\varepsilon dc_{Ni}.
\end{eqnarray}
According to \cite{chap90}, a solution for the function $\xi$ is searched for in the form of the raw of
orthogonal polynomials.
 Sonyne polynomial that were used in the classical work ~\cite{chap90}
are coefficients of the expansion of
the function \\ $(1-s)^{- \frac{3}{2} -1} e^{\frac{xs}{1-s}}$ in powers
of $s$:

\begin{equation}
(1-s)^{- \frac{3}{2} -1} e^{\frac{xs}{1-s}} = \Sigma S^{(p)}_{3/2}(x)s^{p}.
\end{equation}
 Sonyne polynomials are orthogonal:

 \begin{equation}
 \int_0^\infty e^{-x}S^{(p)}_{3/2}(x)S^{(q)}_{3/2}(x)x^{3/2}dx = \frac{\Gamma(p+ \frac{5}{2})}{p!}\delta_{pq},
 \end{equation}
 and

 \begin{eqnarray}
 \begin{aligned}
S^{(0)}_{3/2}(x)  = 1, \quad   S^{(1)}_{3/2}(x)  = \frac{5}{2} - x,\quad
\end{aligned}
\end{eqnarray}
$$
S^{(2)}_{3/2}(x)  = \frac{35}{8} - \frac{7}{2}x+\frac{1}{2}x^{2}.
$$
For a degenerate case we have to seek a solution of ~(\ref{eq_sys}) in
the form of an expansion in polynomials $Q_{n}$  that are orthogonal
with the weight $f_{0}(1-f_{0})x^{3/2}$, analogous to
Sonyne polynomials~\cite{bkrom82}.

\begin{eqnarray}
 \label{sonine}
\begin{aligned}
Q_{0}(x) & = 1, \quad   Q_{1}(x)     = \frac{5G_{5/2}}{2G_{3/2}} - x,
\\
 Q_{2}(x)&  = \frac{35}{8}\frac{G_{7/2}}{G_{3/2}} - \frac{7G_{7/2}}{2G_{5/2}}x+\frac{1}{2}x^{2}, \quad x  = u^{2}.
\end{aligned}
\end{eqnarray}
The nonzero integrals of products of these polynomials with the corresponding weight function are

\begin{widetext}
\begin{eqnarray}
\label{norma}
 \int_0^\infty f_0(1-f_0)\,x^{3/2}Q_{0}^2(x)dx =
 \frac{3}{2}\Gamma(3/2)G_{3/2}(x_0),\qquad \nonumber \qquad\qquad\qquad\\
\int_0^\infty f_0(1-f_0)\,x^{3/2}Q_{1}^2(x)dx
=\frac{15}{4}\Gamma(3/2)G_{3/2}(x_0)\left(\frac{7}{2}\frac{G_{7/2}}{G_{5/2}}-\frac{5}{2}\frac{G_{5/2}^2}{G_{3/2}^2}\right),\qquad\qquad\qquad\\
\int_0^\infty
f_0(1-f_0)\,x^{3/2}Q_{2}^2(x)dx=\frac{105}{16}\Gamma(3/2)G_{3/2}(x_0)
\left(-\frac{35}{8}\frac{G_{7/2}^2}{G_{3/2}^2}+\frac{49}{2}\frac{G_{7/2}^3}{G_{5/2}^2\,G_{3/2}}
-\frac{63}{2}\frac{G_{9/2}G_{7/2}}{G_{5/2}G_{3/2}}+\frac{99}{8}\frac{G_{11/2}}{G_{3/2}^2}\right).\nonumber
\end{eqnarray}
\end{widetext}
We seek $\xi$ and $A_3$ in the form:

\begin{eqnarray}
 \label{polin for A_D}
\begin{aligned}
\xi   = a_{0}Q_{0}+a_{1}Q_{1}+a_{2}Q_{2},\\
A^{(3)}   = c_{0}Q_{0}+c_{1}Q_{1}+c_{2}Q_{2}.
\end{aligned}
\end{eqnarray}
It is easy to show, with account of (\ref{fermiintegral}), (\ref{normalization}), that the first
 term in the right side of (\ref{eq_sys}) is $\sim \frac{m_e}{m_N}$ times
smaller than the second one. Neglecting this term,
 multiplying ~(\ref{eq_sys}) by $RQ_{0}(x)u_{i}$, $RQ_{1}(x)u_{i}$ and $RQ_{2}(x)u_{i}$ and
integrating with respect to $dc_{i}$, we obtain a system of equation
for the heat conductivity coefficients in the form

\begin{widetext}
 \begin{equation}
  \label{general system}
\left\lbrace
\begin{aligned}
0  =-\frac{3}{2} i \omega n_{e}a_{0}+  a_{0}(a_{00}+b_{00}) +a_{1}(a_{01}+b_{01})+a_{2}(a_{02}+b_{02})\\
-\frac{15}{4} n_{e}\left( \frac{7G_{7/2}}{2G_{3/2}} -
\frac{5G_{5/2}^{2}}{2G_{3/2}^{2}}\right)   =- \frac{15}{4} \left(
\frac{7G_{7/2}}{2G_{3/2}} - \frac{5G_{5/2}^{2}}{2G_{3/2}^{2}}\right)
i \omega
n_{e}a_{1} + a_{0}(a_{10}+b_{10})+ a_{1}(a_{11} + b_{11})+a_{2}(a_{12}+b_{12})\\
0 = -
\frac{105}{16}\left(-\frac{35}{8}\frac{G_{7/2}^2}{G_{3/2}^2}+\frac{49}{2}\frac{G_{7/2}^2}{G_{5/2}^2}\frac{G_{7/2}}{G_{3/2}}
-\frac{63}{2}\frac{G_{9/2}G_{7/2}}{G_{5/2}G_{3/2}}+\frac{99}{8}\frac{G_{11/2}}{G_{3/2}}\right)
 i \omega n_{e}a_{2}\qquad\qquad
\end{aligned} \right .
 \end{equation}
 \vspace{-1cm}
 $$+a_{0}(a_{20}+b_{20})+a_{1}(a_{21}+b_{21})+a_{2}(a_{22}+b_{22})$$
 \end{widetext}
 Here $a_{jk}$,  $b_{jk}$ are matrix elements for collision integrals, $\omega = \frac{eB}{m_{e}c}$ is a cyclotron frequency.

     \section{Matrix elements: $b_{jk}$}

 The matrix elements,  $b_{jk}$, connected with electron-nuclei collisions, are determined as follows:

 \begin{eqnarray}
 \label{bjk}
 b_{jk} = R \int f_{0} f_{N0} (1-f_{0}^{'})Q_{j}(u^{2})u_{i}[Q_{k}(u^{2})u_{i} \nonumber\\
 -Q_{k}(u^{'2})u_{i}^{'}]g_{eN}bdb d\varepsilon dc_{Ni} dc_{i}, \\ \nonumber  k\geq 0 .
 \end{eqnarray}

Introduce  functions $\bar\Omega^{(l)}_{eN}(r)$,  defined as (see \cite{chap90})

\begin{equation}
\label{omega0}
\bar\Omega^{(l)}_{eN}(r) = 2\int_0^\infty  f_{0}(1-f_{0})z^{2r+2}\int_0^\infty (1-\cos ^{l} \theta_{12})g_{12} bdbdz,
\end{equation}
where

\begin{equation}
\label{omega1}
z=\left[\frac{m_1 m_2}{2kT(m_1+m_2)}\right]^{1/2}g_{12},\quad g_{12}=|v_1-v_2|,
\end{equation}
for colliding particles "1", "2", $\theta_{12}$ is the scattering angle.
At collisions of electrons ("2") and nuclei ("1")
a mass of the nuclei is much greater than electron
mass $m_{N} \gg m_{e}$, so we can neglect energy exchange in a collision

\begin{equation}
u^{'2}  \approx u^{2},
 \quad  u u^{'} \approx u^{2}(1-cos \theta_{12}),\quad g_{eN}\approx v,\quad z\approx u,
\end{equation}
 Using the relation
~(\ref{nuclear_norm}), we obtain from ~(\ref{bjk})

\begin{widetext}
\begin{equation}
 \label{bjk1}
 b_{jk} = 8\pi^{2}\left( \frac{2kTm_{e}}{h^{2}}\right) ^{3/2}
 \left( \frac{2kT}{m_{e}}\right) ^{1/2}n_{N} \int_0^\infty f_{0}
 (1-f_{0})Q_{j}(x)Q_{k}(x)x^{2}\int_0^\infty(1-\cos \theta_{12} ) bdb  dx,
 \end{equation}
 \end{widetext}
Instead of functions $\bar\Omega^{(l)}_{eN}(r)$ from (\ref{omega0}) we can use in this case
functions $\widehat\Omega^{(l)}_{eN}(r)$ defined as

\begin{equation}
\label{omega}
\widehat\Omega^{(l)}_{eN}(r) = \int_0^\infty  f_{0}(1-f_{0})x^{r+1}\int_0^\infty (1-\cos^{l}\theta_{12}) bdbdx .
\end{equation}
With account of (\ref{sonine}), the elements of a symmetric matrix   $b_{ij}$  are written as

\begin{equation}
\label{b00d}
b_{00} = 8\pi^{2}
 \frac{(2kT)^2}{h^{3}}m_{e} n_{N}
  \widehat\Omega^{(1)}_{eN}(1),
\end{equation}

\begin{widetext}
\begin{equation}
 \label{b01d}
\begin{aligned}
b_{01} =   8\pi^{2}
 \frac{(2kT)^2}{h^{3}}m_{e} n_{N}
\left( \frac{5}{2}
\frac{G_{5/2}}{G_{3/2}}\widehat\Omega^{(1)}_{eN}(1)  - \widehat\Omega^{(1)}_{eN}(2) \right),
\end{aligned}
\end{equation}

\begin{eqnarray}
 \label{b11d}
\begin{aligned}
b_{11}  = 8\pi^{2}
 \frac{(2kT)^2}{h^{3}}m_{e} n_{N}  \left( \frac{25}{4}
\frac{G^{2}_{5/2}}{G^{2}_{3/2}} \widehat\Omega^{(1)}_{eN}(1) -5 \frac{G_{5/2}}{G_{3/2}}
\widehat\Omega^{(1)}_{eN}(2) + \widehat\Omega^{(1)}_{eN}(3)\right),
\end{aligned}
\end{eqnarray}

\vspace{1cm}
\begin{equation}
\begin{aligned}
 \label{b02d}
b_{02} = 8\pi^{2}
\frac{(2kT)^2}{h^{3}}m_{e} n_{N}
 \left[ \frac{35}{8}\frac{G_{7/2}}{G_{3/2}}
  \widehat\Omega^{(1)}_{eN}(1)-\frac{7}{2} \frac{G_{7/2}}{G_{5/2}} \widehat\Omega^{(1)}_{eN}(2) + \frac{1}{2}\widehat\Omega^{(1)}_{eN}(3)\right],
\end{aligned}
\end{equation}

\begin{eqnarray}
 \label{b21d}
\begin{aligned}
b_{12}  = 8\pi^{2}
\frac{(2kT)^2}{h^{3}}m_{e} n_{N}
\left[ \frac{175}{16} \frac{G_{7/2}G_{5/2}}{G^2_{3/2}}\widehat\Omega^{(1)}_{eN}(1)
 -  \frac{105}{8}\frac{G_{7/2}}{G_{3/2}}\widehat\Omega^{(1)}_{eN}(2) + \left(\frac{5}{4}\frac{G_{5/2}}{G_{3/2}}+\frac{7}{2}\frac{G_{7/2}}{G_{5/2}}\right)\widehat\Omega^{(1)}_{eN}(3) - \frac{1}{2}\widehat\Omega^{(1)}_{eN}(4)\right],\,\,\,
\end{aligned}
\end{eqnarray}

\begin{eqnarray}
 \label{b22d}
\begin{aligned}
b_{22}  = 8\pi^{2}
\frac{(2kT)^2}{h^{3}}m_{e} n_{N}
\qquad\qquad\qquad \\
\times \left[\frac{35^2}{8^2}\frac{G^2_{7/2}}{G^2_{3/2}}\widehat\Omega^{(1)}_{eN}(1)
- \frac{245}{8}\frac{G^{2}_{7/2}}{G_{5/2}G_{3/2}}\widehat\Omega^{(1)}_{eN}(2) +(\frac{49}{4}\frac{G_{7/2}^2}{G_{5/2}^2}
+\frac{35}{8} \frac{G_{7/2}}{G_{3/2}})\widehat\Omega^{(1)}_{eN}(3)
-  \frac{7}{2}\frac{G_{7/2}}{G_{5/2}}\widehat\Omega^{(1)}_{eN}(4)+
\frac{1}{4}\widehat\Omega^{(1)}_{eN}(5)\right].
\end{aligned}
\end{eqnarray}
\end{widetext}

\subsection{Functions $\phi^{(l)}_{12}$, and Coulomb logarithm}

The functions $\bar\Omega^{(l)}_{eN}(r)$ from (\ref{omega0}) may be written in the form

\begin{equation}
\label{omegap}
\widehat\Omega^{(l)}_{eN}(r) = 2\left[\frac{m_1 m_2}{2kT(m_1+m_2)}\right]^{1/2}
\end{equation}
$$
\times \int_0^\infty  f_{0}(1-f_{0})z^{2r+2}\phi_{12}^{(l)}dz,
$$
with

\begin{equation}
\label{omegap1}
\phi_{12}^{(l)}=\int_0^\infty (1-\cos ^{l} \theta_{12})g_{12} b \,db.
\end{equation}
During the integration in (\ref{omegap1}) over the impact parameter $db$ the integral has
a logarithmic divergency at infinity. It is removed in a more precise treatment of Coulomb collisions in plasma with account of
correlation functions \cite{balesku}, where the upper limit of the  integration $b_{max}$ appeared.
Introducing $v_0=bg_{12}^2(m_1 m_2/m_0 e_1 e_2)$, where $m_0=m_1+m_2$, $e_1$, $e_2$ are the absolute values of charges,
we obtain after integration \cite{chap90}

\begin{equation}
\label{omegap2}
\phi_{12}^{(1)}=\left(\frac{m_0 e_1 e_2}{m_1 m_2}\right)^2 g_{12}^{-3}\ln(1+v_{0max}^2),
\end{equation}

\begin{equation}
\label{omegap3}
\phi_{12}^{(2)}=2\left(\frac{m_0 e_1 e_2}{m_1 m_2}\right)^2 g_{12}^{-3}\left[\ln(1+v_{0max}^2)-\frac{v^2_{0max}}{1+v^2_{0max}}\right],,
\end{equation}

\begin{equation}
\label{omegap4}
v_{0max}=b_{max} g_{12}^2(m_1 m_2/m_0 e_1 e_2).
\end{equation}
In farther integration the value under the logarithm is taken as constant when the average value $\bar g_{12}$ is taken
instead of the variable $g_{12}$. For the electron-nuclei collisions with $g_{12}\approx v_e$
the approximate expression of the Coulomb logarithm is written in the form \cite{bkphys}

\begin{equation}
 \label{omegap5}
  \Lambda=\frac{1}{2}\ln(1+v_{0max}^2)\approx {\bar \Lambda}_v=\ln\left(b_{\rm max}
  \bar {v_e^2} m_e\over Z{\rm e}^2\right),\quad \Lambda \gg 1,
\end {equation}
where

\begin{eqnarray}
\bar {v_e^2}={3kT\over m_e}\frac{G_{5/2}}{G_{3/2}}=
  {3kT\over m_e} \qquad (ND)  \\ \nonumber
  ={3\over 5}{h^2\over m_{\rm e}^2}\left(3n_e\over 8\pi\right)^{2/3} \qquad (D).
\end{eqnarray}
The value of $b_{max}$ is represented by the radius of Debye screening by electrons $r_{{\cal D}e}$,
and ions $r_{{\cal D}i}$, and may be written as \cite{bkphys}

\begin{equation}
 {1\over {b_{\rm max}}^2}={1\over r_{{\cal D} i}^2}+
  {1\over r_{{\cal D} e}^2}=
  {4\pi{\rm e}^2\over kT}\left(n_NZ^2+n_e \frac{G_{1/2}}{G_{3/2}}\right),
 \end{equation}
  where

\begin{equation}
 \frac{G_{1/2}}{G_{3/2}}=1 \hspace{1cm} (ND)
 \end{equation}
 $$
    \hspace{3cm}  = 4(3\pi^2)^{1/3} \frac{m_e kT}{h^2 n_e^{2/3}}  \qquad (D).
$$
 Influence of quantum effects on the Debye screening was discussed in \cite{yau80}.
The average frequency of  electron-ion collisions $\nu_{ei}$ is written in  \cite{gr70} in the form

\begin{equation}
\label{nuei}
{\nu_{ei}} =  \frac{4}{3 } \sqrt{\frac{2\pi}{m_{e}}}\frac{Z^2e^4n_{N}\Lambda}{(kT)^{3/2}G_{3/2}} \frac{1}{1+e^{-x_0}}.
\end{equation}
In the limiting cases it is expressed as

\begin{equation}
\label{nuen}
 \nu_{ei}={4\over 3}\sqrt{2\pi\over m_e}{Z^2{\rm e}^4n_N\Lambda
      \over (kT)^{3/2}} \qquad (ND)
\end{equation}
 $$
      ={32\pi^2\over 3}m_e{Z^2{\rm e}^4 \Lambda n_N\over h^3 n_{\rm e}}
      \qquad (D).
$$
The average time $\tau_{ei}$ between (ei) collisions is the inverse value of $\nu_{ei}$, and is written as

\begin{equation}
\label{taund}
\tau_{nd} =\frac{1}{{\nu_{nd}}}= {3\over 4}\sqrt{m_e\over 2\pi}{(kT)^{3/2} \over  Z^2{\rm e}^4n_N\Lambda},
\end{equation}

\begin{equation}
\label{taud}
\tau_{d} = \frac{1}{{\nu_{d}}}=\frac{3 h^3 n_{e}}{32 \pi^2 m_{e} Z^2 e^4 \Lambda n_{N}}.
\end{equation}

\subsection{$b_{jk}$ for non-degenerate electrons}

For non-degenerate electrons we have $\exp{(x-x_0)} \ll 1$, $G_n\approx e^{x_0}$, so

\begin{equation}
\widehat\Omega^{(l)}_{eN}(r) = e^{x_0}\int_0^\infty  e^{-x}x^{r+1}\int_0^\infty (1-\cos ^{l} \theta) bdbdx .
\end{equation}
 In this case it is more convenient to use functions
 $\Omega^{(l)}_{eN}(r) $  defined as

\begin{equation}
\label{omegafor nondeg}
\Omega^{(l)}_{eN}(r) = \frac{\sqrt\pi}{2} \int_{0}^{\infty}  e^{-x}x^{r+1}
\int_{0}^{\infty} (1-\cos ^{l} \theta) bdbdx.
\end{equation}
Using (\ref{omegap}),(\ref{omegap2}) we find expressions for non-degenerate case as

\begin{equation}
\label{omeganondeg}
\Omega^{(1)}_{eN}(r) = \sqrt\pi \frac{e^{4}\Lambda Z^2}{(2kT)^2}\Gamma(r),\quad \Gamma(1)=1;
\end{equation}
$$
\Gamma(2)=1; \quad \Gamma(3)=2; \quad \Gamma(4)=6; \quad \Gamma(5)=24.
$$
Substituting (\ref{omeganondeg}) into (\ref{b00d})-(\ref{b22d}), taking into account that

\begin{equation}
\label{x0nd}
G_n=e^{x_0},\qquad
e^{x_{0}} = \frac{n_{e}}{2\pi^{3/2}}\left(\frac{h^2}{2kTm_{e}}\right)^{3/2},
\end{equation}
and using (\ref{nuen}), (\ref{taund}), we write $b_{jk}$ for non-degenerate electrons as

\begin{equation}
\label{b00nd}
b_{00} =8\sqrt{\pi}
\frac{n_{e} n_{N} e^4 Z^2\Lambda}{(2kT)^{3/2}\sqrt{m_{e}}}=\frac{3n_e}{2\tau_{nd}},
\end{equation}

\begin{equation}
\label{b01nd}
b_{01}=12\sqrt{\pi}
\frac{n_{e} n_{N} e^4 Z^2\Lambda}{(2kT)^{3/2}\sqrt{m_{e}}}=\frac{9n_e}{4\tau_{nd}},
\end{equation}
\vspace{0.15cm}

\begin{eqnarray}
\label{b11nd}
b_{11}  = 26\sqrt{\pi}
\frac{n_{e} n_{N} e^4 Z^2\Lambda}{(2kT)^{3/2}\sqrt{m_{e}}}=\frac{39n_e}{8\tau_{nd}},
\end{eqnarray}

\begin{equation}
\label{b02nd}
\begin{aligned}
b_{02} = 15\sqrt{\pi}
\frac{n_{e} n_{N} e^4 Z^2\Lambda}{(2kT)^{3/2}\sqrt{m_{e}}}=\frac{45n_e}{16\tau_{nd}},
\end{aligned}
\end{equation}

\begin{equation}
\label{b12nd}
b_{12} = \frac{69\sqrt{\pi}}{2}
\frac{n_{e} n_{N} e^4 Z^2\Lambda}{(2kT)^{3/2}\sqrt{m_{e}}}=\frac{207n_e}{32\tau_{nd}},
\end{equation}

\begin{equation}
\label{b22nd}
b_{22} = \frac{433\sqrt{\pi}}{8}
\frac{n_{e} n_{N} e^4 Z^2\Lambda}{(2kT)^{3/2}\sqrt{m_{e}}}=\frac{1299n_e}{128\tau_{nd}}.
\end{equation}

\subsection{$b_{jk}$ for partially degenerate electrons}

To calculate $b_{ij}$ for degenerate electrons we  use expressions for $\widehat\Omega^{(1)}_{eN}(r)$ and  $ G_{n}(x_0)$.
With account of (\ref{omegap}),(\ref{omegap2}) we obtain

\begin{widetext}
\begin{equation}
\label{omegadeg}
\widehat\Omega^{(1)}_{eN}(r) = \int_0^\infty f_{0}(1-f_{0})x^{r+1} \int_0^{b_{max}} (1-\cos\theta) b db dx =
2\frac{e^{4} Z^2 \Lambda}{(2kT)^2}\int_0^\infty f_{0}(1-f_{0})x^{r-1}dx=2\frac{e^{4} Z^2 \Lambda}{(2kT)^2}\Gamma(r)G_{r-1}(x_0).
\end{equation}
\end{widetext}
The integral in (\ref{omegadeg}) is calculated exactly for $r=1$
$\int_0^\infty f_{0}(1-f_{0})x^{r-1}dx=[1+\exp(-x_0)]^{-1}$.
Between non-degenerate electrons with large negative non-dimensional
chemical potential at $x_0 \ll -1$, and strongly degenerate electrons
with $x_0 \gg 1$ there is a level of degeneracy at which $x_0=0$.
   Let us calculate matrix elements $b_{jk}$ for the level of degeneracy,
corresponding to $x_0=0$.
The functions $G_n(0)$ have the following
numerical values, according to \cite{dau90}

\begin{equation}
 \label{x00}
G_{n}(0)=\frac 1{\Gamma(n)}\int_0^\infty\frac{x^{n-1}dx}{1+e^x} =
(1-2^{1-n})\zeta(n),
 \end{equation}
where Riemann $\zeta$-function has the following values for the
indexes used here \cite{ye}

$$\zeta(3/2)=2.612,\quad \zeta(5/2)=1.341,\quad \zeta((7/2)=1.127,$$
\begin{equation}
 \label{xi}
\zeta(9/2)=1.0547,\quad \zeta(11/2)=1.0252,\quad \zeta(2)=1.645,
\end{equation}
$$ \zeta(3)=1.202,\quad \zeta(4)=1.0823, \quad \zeta(5)=1.0369.$$
The functions $G_{n}(0)$ have the following values

$$G_{3/2}(0)=0.765,\quad G_{5/2}(0)=0.867,\quad G_{7/2}(0)=0.928,$$
\begin{equation}
 \label{g0}
G_{9/2}(0)=0.9615,\quad G_{11/2}=0.980,\quad G_{2}(0)=1.645,
\end{equation}
$$G_{3}(0)=0.9015,\quad G_{4}(0)=0.947,\quad G_{5}(0)=0.972.$$
The functions $\widehat\Omega^{(1)}_{eN}(r)$ at $x_0=0$, defined as $\widehat\Omega^{(1)}_{eN0}(r)$,
have the following values, according to (\ref{omegadeg}), using (\ref{x00})-(\ref{g0})

$$
\widehat\Omega^{(1)}_{eN0}(1)=\frac{e^4Z^2\Lambda}{(2kT)^2}\equiv\, I,
\quad \widehat\Omega^{(1)}_{eN0}(2)=2\ln{2}\,I=1.39\,I,
$$
\begin{equation}
 \label{omega00}
\widehat\Omega^{(1)}_{eN0}(3)=3.29\,I,\quad
\widehat\Omega^{(1)}_{eN0}(4)=10.82\,I,\quad
\end{equation}
$$\widehat\Omega^{(1)}_{eN0}(5)=45.46\,I.
$$
The level of degeneracy $DL(0)$ at $x_0=0$ is defined as a ratio of
the Fermi energy $\varepsilon_{fe}$ to $kT$. With account of
(\ref{el_num-den}), (\ref{g0}), we obtain
\begin{equation}
 \label{deg0}
DL(0)=\frac{\varepsilon_{fe}}{kT}=\frac{(3\pi^2
n_e)^{2/3}h^2}{8\pi^2 m_e kT}
\end{equation}
$$=\frac{\pi}{4}\left(\frac{3}{\pi}\right)^{2/3}[2G_{3/2}(0)]^{2/3}=1.011.
$$
At $x_0=0$ the expression for the electron concentration $n_{e0}$ from (\ref{el_num-den}), and
the average time $\tau_{ei}$ between (ei) collisions, which is the inverse value of $\nu_{ei}$,
are written, using (\ref{nuei}),(\ref{g0}),(\ref{fermiintegral}) as

\begin{equation}
\label{nex0}
n_{e0}=2G_{3/2}(0) \left( \frac{\mathit{kTm}_e}{2\pi\hbar^2}\right) ^{3/2}
=2\times 0.765  \left( \frac{\mathit{kTm}_e}{2\pi\hbar^2}\right) ^{3/2},
\end{equation}
\begin{equation}
\label{tauei}
{\tau_{d0}} =  \frac{3}{4 } \sqrt{\frac{m_{e}}{2\pi}}\frac{(kT)^{3/2}G_{3/2}}{Z^2e^4n_{N}\Lambda} (1+e^{-x_0})
\end{equation}
$$=0.765  \frac{3}{2} \sqrt{\frac{m_{e}}{2\pi}}\frac{(kT)^{3/2}}{Z^2e^4n_{N}\Lambda}.
$$
Using (\ref{omega00}), (\ref{nex0}, (\ref{tauei}) we find from (\ref{b00d})-(\ref{b22d})

\begin{equation}
\label{b00d0} b_{00} = \frac{8\pi^2e^4 Z^2 m_e\Lambda}{h^3}n_N=\frac{3}{2}\frac{n_{e0}}{\tau_{d0}}
\end{equation}

\begin{equation}
 \label{b01d0}
\begin{aligned}
b_{01} = 2.16\frac{n_{e0}}{\tau_{d0}},\quad b_{11} = 5.162\frac{n_{e0}}{\tau_{d0}},\quad b_{02} =
2.588\frac{n_{e0}}{\tau_{d0}},
\end{aligned}
\end{equation}

\begin{eqnarray}
 \label{b22d0}
\begin{aligned}
b_{12}  = 6.671\frac{n_{e0}}{\tau_{d0}},\quad b_{22} = 11.038\frac{n_{e0}}{\tau_{d0}}.
\end{aligned}
\end{eqnarray}
For arbitrary level of degeneracy at $x_0\neq 0$ the functions $G_n(x_0)$ in (\ref{fermiintegral})
are not expressed  analytically, and should be calculated numerically for each $x_0$, at corresponding
DL

\begin{equation}
 \label{degx0}
DL(x_0)=\frac{(3\pi^2 n_e)^{2/3}h^2}{8\pi^2 m_e kT}
=\frac{\pi}{4}\left(\frac{3}{\pi}\right)^{2/3}[2G_{3/2}(x_0)]^{2/3}.
\end{equation}
After numerical calculation of $G_n(x_0)$, the matrix elements $b_{jk}$ at arbitrary $x_0$ are found in the
same way as it is done above at $x_0=0$.


\subsection{$b_{jk}$ for strongly degenerate electrons}

For strongly degenerate case $x_0 \gg 1$ we use  \cite{dau90} the following expansions
\begin{widetext}
\begin{equation}
\begin{aligned}
 \label{gammaexp}
 G_{r}(x_0)  = \frac{1}{\Gamma (r)}\left[ \frac{x_{0}^{r}}{r} +
 \frac{\pi^{2}}{6}(r-1)x^{r-2}_{0}
 +\frac{7\pi^4}{360}(r-1)(r-2)(r-3)x_0^{r-4}\right]\,\,\,{\rm for}\,\, r\ge
 1.
 \end{aligned}
\end{equation}

\begin{equation}
\begin{aligned}
 \label{gammaexpansion}
 \Gamma(r)G_{r-1}(x_0)  = (r-1)\left[ \frac{x_{0}^{r-1}}{r-1} +
 \frac{\pi^{2}}{6}(r-2)x^{r-3}_{0}
 +\frac{7\pi^4}{360}(r-2)(r-3)(r-4)x_0^{r-5}\right]\,\,\,{\rm for}\,\, r\ge
 2.
 \end{aligned}
\end{equation}
\end{widetext}
 For strongly degenerate electrons

\begin{equation}
 \label{x0}
x_0 =\frac{(3\pi^2 n_e)^{2/3}h^2}{8\pi^2 m_e kT} \gg 1.
\end{equation}
We obtain than from (\ref{omegadeg}),(\ref{gammaexpansion}), omitting  exponentially small terms $\sim e^{-x_0}$

\begin{equation}
 \label{odeg1}
\widehat\Omega^{(1)}_{eN}(1) = 2\frac{e^{4} Z^2 \Lambda}{(2kT)^2}\int_0^\infty f_{0}(1-f_{0})dx=2\frac{e^{4} Z^2 \Lambda}{(2kT)^2},
\end{equation}
\begin{equation}
 \label{odeg2}
\widehat\Omega^{(1)}_{eN}(2) = 2\frac{e^{4} Z^2 \Lambda}{(2kT)^2}\Gamma(2)G_1(x_0)=2\frac{e^{4} Z^2 \Lambda}{(2kT)^2} x_0,
\end{equation}

\begin{widetext}
\begin{equation}
 \label{odeg3}
\widehat\Omega^{(1)}_{eN}(3) = 2\frac{e^{4} Z^2 \Lambda}{(2kT)^2}\Gamma(3)G_2(x_0)=2\frac{e^{4} Z^2 \Lambda}{(2kT)^2}
(x_0^2+\frac{\pi^2}{3}),
\end{equation}
 \begin{equation}
 \label{odeg4}
\widehat\Omega^{(1)}_{eN}(4) = 2\frac{e^{4} Z^2 \Lambda}{(2kT)^2}\Gamma(4)G_3(x_0)=2\frac{e^{4} Z^2 \Lambda}{(2kT)^2}
(x_0^3+\pi^2 x_0),
\end{equation}
\begin{equation}
 \label{odeg5}
\widehat\Omega^{(1)}_{eN}(5) = 2\frac{e^{4} Z^2 \Lambda}{(2kT)^2}\Gamma(5)G_4(x_0)=2\frac{e^{4} Z^2 \Lambda}{(2kT)^2}
 (x_o^4+2\pi^2x_0^2+\frac{7\pi^2}{15}).
\end{equation}
\end{widetext}
Using (\ref{gammaexp}),(\ref{odeg1})-(\ref{odeg5}), we find from (\ref{b00d})-(\ref{b22d})

\begin{equation}
\label{b00ds}
b_{00} = \frac{16\pi^2e^4 Z^2 m_e\Lambda}{h^3}n_N=\frac{3n_e}{2\tau_{d}},
\end{equation}

\begin{equation}
 \label{b01ds}
\begin{aligned}
b_{01} = \frac{3\pi^2n_e}{4x_0\tau_{d}},\quad b_{11} = \frac{\pi^2
n_e}{2\tau_{d}},\quad b_{02} = -\frac{7\pi^4 n_e}{320
x_0^2\tau_{d}},
\end{aligned}
\end{equation}

\begin{eqnarray}
 \label{b22ds}
\begin{aligned}
b_{12}=\frac{709\pi^4 n_e}{960x_0\tau_{d}}, \quad b_{22}=\frac{2
\pi^4n_e}{15\tau_{d}}.
\end{aligned}
\end{eqnarray}
\vspace{1cm}
\section{Matrix elements: $a_{jk}$}

 The matrix elements, $ a_{jk}$, related to electron-electron collisions, are determined as follows:

 \begin{widetext}
 \begin{equation}
 \label{aji}
a_{jk}  = R^{2} \int  f_{0}f_{01}(1 - f_{0}^{'})(1- f_{01}^{'})Q_{j}
(u^{2})u_{i}[Q_{k}(u^{2})u_{i}  + Q_{k}(u_{1}^{2})u_{1i}- Q_{k}(u^{'2})u_{i}^{'} - Q_{k}(u_{1}^{'2})u_{1i}^{'}]g_{ee}bdb d\varepsilon dc_{1i}dc_{i},
\end{equation}
\end{widetext}
Let's introduce following variables \cite{chap90}
\begin{equation}
  \label{aj2}
 \begin{aligned}
 G_{li}  = \frac{1}{2}(c_{i}+c_{1i})=\frac{1}{2}(c^{'}_{i}+c^{'}_{1i}),\qquad\\
  g_{ee,i} = c_{1i}-c_{i},\quad g_{ee,i}^{'} = c_{1i}^{'}-c_{i}^{'},\qquad\\
  g_{ee}=|g_{ee,i}|=|g_{ee,i}^{'}|=g_{ee}^{'},\quad G_{0i}=G_{li}-c_{0i},\\
  v_{i}=G_{0i}-\frac{1}{2}g_{ee,i},\quad v_{i1}=G_{0i}+\frac{1}{2}g_{ee,i},\quad\\
  v^2+v_1^2=2G_0^2+\frac{1}{2}g_{ee}^2\quad \qquad.
 \end{aligned}
 \end{equation}
Here $G_{li}$ is a velocity of the center of mass of two colliding electrons in the laboratory frame,
$G_{0i}$ is the same value in the comoving frame, $g_{ee,i}$ is a relative velocity of two colliding electrons
before encounter, $g_{ee,i}^{'}$ is the same value after encounter; $v_i$ and $v_{1i}$ are  velocities of
colliding electrons in the comoving frame, defined above. Introduce non-dimensional variables

 \begin{equation}
  \label{aj3}
 \begin{aligned}
  g_{i}  = \frac{1}{2}\left( \frac{m_{e}}{kT}\right)^{1/2} g_{ee,i},
  \quad  g_{i}^{'}  = \frac{1}{2}\left( \frac{m_{e}}{kT}\right)^{1/2} g_{ee,i}^{'},\qquad\\
g = |g_i|=|g^{'}_i|=g^{'},\qquad G_i=\left(\frac{m_{e}}{kT}\right)^{1/2}G_{0i},\qquad\\
 dc_{i}dc_{1i}  = \left( \frac{2kT}{m_{e}}\right) ^{3}dG_{i}dg_{i},\quad\qquad\\
 u^2+u_1^2=G^2+g^2,\quad u^2=u_i^2, \quad u_{1}^2=u_{1i}^2,\quad G^2=G_i^2.
 \end{aligned}
 \end{equation}
Here $u_i$, $u_{1i}$ are non-dimensional velocities of electrons, defined above.  The matrix elements

\begin{widetext}
\begin{equation}
 \label{aj4}
\begin{aligned}
a_{j0}  = 8 \left( \frac{2kTm_{e}}{h^{2}}\right)^{3}
\left( \frac{kT}{m_{e}}\right)^{1/2}\int f_{0}f_{01}(1-f^{'}_{01})(1-f^{'}_{0}) Q_{j}(u^2) u_{i}  [u_{i}+ u_{1i}  - u^{'}_{i}- u_{1i}^{'}]gbdbd\varepsilon dg_idG_i=0.
\end{aligned}
\end{equation}
\end{widetext}
$a_{j0}$ are equal to zero because the momentum conservation during encounter define the zero value in the brackets of (\ref{aj4}).
The nonzero matrix elements $a_{jk}\,\,(j,k\ge 1)$ are defined as

\begin{widetext}
\begin{equation}
 \label{a11gen}
\begin{aligned}
a_{jk}  = 8 \left( \frac{2kTm_{e}}{h^{2}}\right)^{3}
\left( \frac{kT}{m_{e}}\right)^{1/2}\int f_{0}f_{01}(1-f^{'}_{01})(1-f^{'}_{0}) Q_{j}u_{i}[Q_{k} u_{i}+Q_{k} u_{1i} -Q_{k}^{'} u^{'}_{i}-Q_{k}^{'} u_{1i}^{'}]gbdbd\varepsilon dg_idG_i.
\end{aligned}
\end{equation}
\end{widetext}
Here $Q_i$ are function of $u^2$ or $u_1^2$, and $Q_i^{'}$ are function of $u^{'2}$ or $u_1^{'2}$ respectively.

\subsection{$a_{jk}$ for non-degenerate electrons}

For non-degenerate case, at $x_0\gg 1$, $f_0 \ll 1$, polynomials $Q_i$ are reduced to $S_{3/2}^{(i)}$, and  we have from (\ref{a11gen}) the following expression $(j,k\ge 1)$

\begin{widetext}
\begin{equation}
 \label{a11nd}
\begin{aligned}
a_{jk}  = 8 \left( \frac{2kTm_{e}}{h^{2}}\right)^{3}
\left( \frac{kT}{m_{e}}\right)^{1/2} e^{2x_{0}}\int e^{-u^2-u^{'2}} S_{3/2}^{(j)}u_{i}[S_{3/2}^{(k)} u_{i}+S_{3/2}^{(k)} u_{1i} -S_{3/2}^{(k)'} u^{'}_{i}-S_{3/2}^{(k)'} u_{1i}^{'}]gbdbd\varepsilon dg_idG_i.
\end{aligned}
\end{equation}
\end{widetext}
The integrals

\begin{widetext}
\begin{equation}
 \label{a11nd1}
[S_{3/2}^{(j)},S_{3/2}^{(k)}]  = \frac{2}{\pi^3}
\left( \frac{kT}{m_{e}}\right)^{1/2} \int e^{-u^2-u^{'2}} S_{3/2}^{(j)}u_{i}[S_{3/2}^{(k)} u_{i}+S_{3/2}^{(k)} u_{1i} -S_{3/2}^{(k)'} u^{'}_{i}-S_{3/2}^{(k)'} u_{1i}^{'}]gbdbd\varepsilon dg_idG_i
\end{equation}
\end{widetext}
are calculated in \cite{chap90}, and are defined by formulae

\begin{equation}
\label{a11nd2}
\begin{aligned}
\quad [S_{3/2}^{(1)},S_{3/2}^{(1)}]= 4\Omega_{ee}^{(2)}(2),\qquad\quad\\
[S_{3/2}^{(1)},S_{3/2}^{(2)}]= 7\Omega_{ee}^{(2)}(2)-2\Omega_{ee}^{(2)}(3),\quad\\
[S_{3/2}^{(2)},S_{3/2}^{(2)}]=\frac{77}{4}\Omega_{ee}^{(2)}(2)-7\Omega_{ee}^{(2)}(3)+\Omega_{ee}^{(2)}(4).
\end{aligned}
\end{equation}
The functions $\Omega_{ee}^{(l)}(r)$ are similar to functions (\ref{omegafor nondeg}), and are defined in \cite{chap90} as

\begin{equation}
\label{a11nd3}
\begin{aligned}
\Omega_{ee}^{(l)}(r) = \frac{\sqrt\pi}{2} \int_{0}^{\infty}  e^{-x}x^{r+\frac{1}{2}}\phi_{ee}(l)dx,\\
\phi_{ee}(l)=\int_{0}^{\infty} (1-\cos ^{l} \theta) bdb,\quad x=g^2.
\end{aligned}
\end{equation}
Using (\ref{a11nd1}),(\ref{a11nd2}),(\ref{x0nd}) in (\ref{a11nd}), we have

\begin{equation}
 \label{a11nd4}
\begin{aligned}
a_{jk}  = n_e^2 [S_{3/2}^{(j)},S_{3/2}^{(k)}]
\end{aligned}
\end{equation}
For plasma with $\Lambda\gg 1$ from (\ref{omegap5}) we have from (\ref{omegap3}), (\ref{a11nd3})

\begin{equation}
 \label{a11nd5}
\begin{aligned}
\phi_{ee}(2)\approx \frac{16e^4}{m_e^2 g_{ee}^3},\qquad\\
\Omega_{ee}^{(2)}(r)=\sqrt{\pi}\frac{e^4 \Lambda}{\sqrt{m_e} (kT)^{3/2}}\Gamma(r).
\end{aligned}
\end{equation}
Using (\ref{a11nd5}), we have from (\ref{a11nd2}),(\ref{a11nd4}), with account of (\ref{taund}), with $n_e= Z n_N$

 \begin{equation}
\label{a11nd0}
a_{11} = 4 n^{2}_{e}\frac{\sqrt{\pi} \Lambda e^4}{\sqrt{m_e} (kT)^{3/2}}=\frac{3}{\sqrt{2}}\frac{n_e}{Z\tau_{nd}},
\end{equation}

\begin{equation}
\label{a12nd}
a_{12} = 3 n^{2}_{e} \frac{\sqrt{\pi} \Lambda e^4}{\sqrt{m_e}(kT)^{3/2}}= \frac{9}{4 \sqrt{2}}\frac{n_e}{Z\tau_{nd}},
\end{equation}

\begin{equation}
\label{a22nd}
a_{22} = \frac{45}{4} n^{2}_{e}\frac{\sqrt{\pi} \Lambda e^4}{\sqrt{m_e}(kT)^{3/2}}=\frac{135}{16 \sqrt{2}}\frac{n_e}{Z\tau_{nd}}.
\end{equation}

\subsection{$a_{jk}$ for degenerate electrons}

The matrix elements $a_{jk}$ for strongly degenerate case had been found analytically in \cite{tom}, see also \cite{bkrom82}.
They were calculated for strongly degenerate neutrons in a nuclear matter in \cite{tom}, and for the neutrons in the inner crust of a neutron star,
with many free neutrons \cite{bkrom82}. It was found in the last case that in presence of nondegenerate heavy nuclei, and strongly degenerate
neutron, the input of collisions between them in the heat transfer and diffusion coefficients is negligibly small, in comparison with neutron-nuclei collisions.
The same situation we have for the strongly degenerate electrons, for which, using results of \cite{tom},
the estimations give $a_{jk}\sim b_{jk}/x_0^2\ll b_{jk}$  for $x_0\gg 1$. Therefore for strongly degenerate
electrons the Lorentz approximation, with account of collisions between light and heavy particles only, is asymptotically exact. The heat transfer
coefficients for strongly degenerate electrons in presence of a magnetic field are calculated in section VIII.

The situation is more complicated for partially degenerate electrons. In this case there are no analytical expressions for the matrix elements $a_{jk}$,
which should be found numerically by integration of multi-dimensional integrals in (\ref{a11gen}). Another problem is more serious. As shown on section VII,
the precision of polynomial approximation is decreasing with increasing of the level of degeneracy. For non-degenerate electrons the result of
three-polynomial approximation in Lorentz gas at $B=0$ is less than the exact result in Lorentz approximation by only about 2.2\%, see (\ref{lam3l}) and (\ref{laml}). Similar
calculations for moderately degenerate electrons at $x_0=0$ in (\ref{qu302}) and (\ref{qu0l}) show that the result of three polynomial approximation is
about 87\% of the exact result. Therefore for stronger degeneracy the result of 3-polynomial approximation will be even farther (less) from the exact result,
and to obtain good results in the polynomial approximation the number of polynomials should increase with the level of degeneracy. That leads to
very cumbersome analytical calculations. In looks out that it is better to solve this problem by numerical calculations, if a good precision is needed.
In  astrophysical problems it could be enough to use the interpolation formulae between sufficiently exact results obtained in 3-polynomial
approximation for nondegenerate electrons
with account of $e\,e$ collisions, and asymptotically exact result for strongly degenerate electrons in Lorentz approximation. The discussion of this problem is
given in section IX.

\section{Tensor of a heat conductivity}

The  heat flux is expressed via the heat conductivity tensor in the form \cite{bk64, bkl88, kl64}:

\begin{equation}
  \label{q}
 q_{i} = -\lambda_{ik} \frac{\partial T}{\partial r_{k}},
 \end{equation}
where $\lambda_{ik}$:

$$
 \lambda_{ik} = \frac{5}{2}\frac{k^{2}Tn_e}{m_e} \frac{G_{5/2}}{G_{3/2}}   \left\{ \left[a_{0}^{1} - \left( \frac{7}{2} \frac{G_{7/2}}{G_{5/2}} - \frac{5}{2} \frac{G_{5/2}}{G_{3/2}}\right) a_{1}^{1} \right]\delta_{ik}\right.
$$
 \begin{eqnarray}
\label{tensor general}
  - \left.\varepsilon_{ikn}B_{n}\left[b_{0}^{1} - \left( \frac{7}{2} \frac{G_{7/2}}{G_{5/2}} - \frac{5}{2} \frac{G_{5/2}}{G_{3/2}}\right) b_{1}^{1}\right] \right.
 \end{eqnarray}
 $$ \left. +B_{i}B_{k}\left[c_{0}^{1} - \left( \frac{7}{2} \frac{G_{7/2}}{G_{5/2}} - \frac{5}{2} \frac{G_{5/2}}{G_{3/2}}\right)c_{1}^{1}\right] \right\}
 $$
Here $a_{0}^1$, $a_{1}^1$; and $b_{0}^1$, $b_{1}^1$ are the real and imaginary parts of the coefficients $a_{0}$ and $a_{1}$, respectively:

\begin{equation}
\label{exp_for_tens}
\begin{aligned}
a_{0}  & = a_{0}^{1}+iBb_{0}^{1}, &
a_{1}  & = a_{1}^{1}+iBb_{1}^{1} \\
B^2\,c_{0}^{1} & = (a_{0}^{1})_{B=0} - a_{0}^{1},
 &  B^2\,c_{1}^{1} & = (a_{1}^{1})_{B=0}-a_{1}^{1}
\end{aligned}
\end{equation}
To find the coefficients  $a_{0}$,$a_{1}$ for arbitrary electron degeneracy it is necessary to
solve the system of equations (\ref{general system}) with matrix elements $b_{jk}$ from (\ref{b00d})-(\ref{b22d}), and
matrix elements $a_{jk}$ from  section V. For arbitrary degeneracy of electrons the coefficients in the heat conductivity
tensor, as well as well as in  3 other tensors defining the transport of a heat and electrical current in a dense plasma, may be evaluated only numerically.
In two limiting cases of non-degenerate, and strongly degenerate electrons the results are found analytically.

\bigskip

\subsection{Heat conductivity tensor for non-degenerate electrons}

For non-degenerate electrons tensor (\ref{tensor general}) can be written as follows:

\begin{widetext}
\begin{eqnarray}
\label{tensor nondeg}
 \lambda_{ik} = \frac{5}{2}\frac{k^{2}Tn_e}{m_e}     \left[(a_{0}^{1} -  a_{1}^{1})\delta_{ik} - \varepsilon_{ikn}B_{n}(b_{0}^{1} - b_{1}^{1})   + B_{i}B_{k}(c_{0}^{1} -
 c_{1}^{1})\right]
 \end{eqnarray}
 \end{widetext}
The system for 3-polynomial solution  for the electrons  in presence of magnetic field, following from (\ref{general system}), with account of (\ref{b00nd})-(\ref{b22nd}), (\ref{a11nd0})-(\ref{a22nd}), is written as

\begin{widetext}
\begin{equation}
 \label{system3nd}
\left\lbrace
\begin{aligned}
0  = -\frac{3}{2} i\omega\tau_{nd} a_{0} +\frac{3}{2}a_{0} +\frac{9}{4}a_{1}+\frac{45}{16}a_{2}\\
-\frac{15}{4} \tau_{nd} =-\frac{15}{4}i\omega\tau_{nd} a_{1} +\frac{9}{4}a_{0}+ \frac{3}{2}\left(\frac{13}{4}
+\frac{\sqrt{2}}{Z}\right) a_{1}+\frac{9}{8}\left(\frac{23}{4}+\frac{\sqrt{2}}{Z}\right) a_{2}\\
0 = -\frac{105}{16} i \omega\tau_{nd} a_{2} + \frac{45}{16}a_{0}+\frac{9}{8}\left(\frac{23}{4}+\frac{\sqrt{2}}{Z}\right)a_{1}
+\frac{3}{32}\left(\frac{433}{4}+\frac{45\sqrt{2}}{Z}\right)a_{2}
\end{aligned} \right .
 \end{equation}
\end{widetext}
Two first equations at $a_2=0$
determine the 2-polynomial approximation, giving with account of (\ref{tensor nondeg}) the following results for the case $B=0$

\begin{equation}
\begin{aligned}
\label{qu2t}
a_0=\frac{15}{4}\frac{\tau_{nd}}{1+\frac{\sqrt{2}}Z},\quad a_1=-\frac{5}{2}\frac{\tau_{nd}}{1+\frac{\sqrt{2}}Z},\qquad\qquad\\
\end{aligned}
 \end{equation}
\begin{equation}
\label{qu2t1}
\lambda^{(2)}_{nd}=\frac{125}{8}\frac{k^2 T n_e}{m_e}\frac{\tau_{nd}}{1+\frac{\sqrt{2}}Z}=15.63\frac{k^2 T n_e}{m_e}\frac{\tau_{nd}}{1+\frac{\sqrt{2}}Z}.
 \end{equation}
The above results coincide with the results obtained in
\cite{marsh90},\cite{bk64}.
  In 3-polynomial approximation we obtain  the solution of
(\ref{system3nd}) for $a_{0},\,a_{1}$, and heat conductivity
coefficient for the case $B=0$,
 with account of (\ref{tensor nondeg}), as

\begin{eqnarray}
\begin{aligned}
\label{qu3t}
 a_0=\frac{165}{32} \frac{1+\frac{15\sqrt{2}}{11Z}}{1+\frac{61\sqrt{2}}{16Z}+\frac{9}{2Z^2}}\tau_{nd},\qquad\qquad\\
 a_1=-\frac{65}{8}\frac{1+\frac{45\sqrt{2}}{52\,Z}}{1+\frac{61\sqrt{2}}{16Z}+\frac{9}{2Z^2}}\tau_{nd},\qquad\qquad\\
\end{aligned}
 \end{eqnarray}


\begin{eqnarray}
\begin{aligned}
\label{qu3t1}
\lambda^{(3)}_{nd}
=\frac{2125}{64}\frac{k^2 T n_e}{m_e}\frac{1+\frac{18\sqrt{2}}{17Z}}{1+\frac{61\sqrt{2}}{16Z}+\frac{9}{2Z^2}}\tau_{nd},\quad\\
=33.20\frac{k^2 T n_e}{m_e}\frac{1+\frac{18\sqrt{2}}{17Z}}{1+\frac{61\sqrt{2}}{16Z}+\frac{9}{2Z^2}}\tau_{nd},\quad.
 \end{aligned}
 \end{eqnarray}
The value

\begin{equation}
 \label{eecoll}
Q=\frac{64 m_e \lambda^{(3)}_{nd}}{2125 k^2 T n_e \tau_{nd}}=\frac{1+\frac{18\sqrt{2}}{17Z}}{1+\frac{61\sqrt{2}}{16Z}+\frac{9}{2Z^2}},
\end{equation}
showing how nondegenerate electron-electron collisions decrease the heat conductivity coefficient at $B=0$ is presented in Table 1 for different values of $Z$.

\begin{table}[h]
\caption{\label{t7}  The values of $Q$ for different elements: hydrogen (Z=1); helium (Z=2); carbon (Z=6); oxygen (Z=8); iron (Z=26), which may be expected
in the outer layers of white dwarfs and neutron stars.}
\centering
\bigskip
\begin{tabular}{lccccccc}
\hline\hline
{\bf Z}\,\,\, &1\,\,\, &2\,\,\, &6\,\,\, &8\,\,\, &26\,\,\, &$\infty$\\
\hline
\\
{\bf Q}\,\,\, &0.23\,\,\, &0.365\,\,\, &0.62\,\,\, &0.68\,\,\, & 0.87\,\,\,&1\\
\hline
\end{tabular}
\end{table}

In two polynomial approximation,
taking $a_2=0$, we obtain the solution of the system (\ref{system3nd}) in the form

\begin{equation}
 \label{a02t1}
a_{0} = \frac{15}{4}\tau_{nd}\frac{1}{1+\frac{\sqrt{2}}{Z}- \frac{5}{2}\omega^2\tau_{nd}^2-\left(\frac{23}{4}+\frac{\sqrt{2}}{Z}\right)i\omega\tau_{nd}},
\end{equation}

 \begin{equation}
 \label{a12t1}
a_{1} = -\frac{5}{2}\tau_{nd}\frac{1-i\omega\tau_{nd} }{1+\frac{\sqrt{2}}{Z}- \frac{5}{2}\omega^2\tau_{nd}^2-\left(\frac{23}{4}+\frac{\sqrt{2}}{Z}\right)i\omega\tau_{nd}},
\end{equation}

\begin{widetext}
\begin{equation}
 \label{a02t}
a_{0}^1 = \frac{15}{4}\tau_{nd}\frac{1+\frac{\sqrt{2}}{Z} - \frac{5}{2} \omega^2 \tau_{nd}^2 }{\left(1+\frac{\sqrt{2}}{Z}\right)^2+ \left(\frac{449}{16}+\frac{13}{2}\frac{\sqrt{2}}{Z}+\frac{2}{Z^2}\right)\omega^2\tau_{nd}^2
+ \frac{25}{4}\omega^4\tau_{nd}^4},
\end{equation}

 \begin{equation}
 \label{b02t}
b_{0}^1 = \frac{15}{4}\frac{\omega\tau_{nd}^2}{B}\frac{\frac{23}{4}+\frac{\sqrt{2}}{Z}}{\left(1+\frac{\sqrt{2}}{Z}\right)^2+ \left(\frac{449}{16}+\frac{13}{2}\frac{\sqrt{2}}{Z}+\frac{2}{Z^2}\right)\omega^2\tau_{nd}^2
+ \frac{25}{4}\omega^4\tau_{nd}^4},
\end{equation}

 \begin{equation}
 \label{a12t}
a_{1}^1 = -\frac{5}{2}\tau_{nd}\frac{1+ \frac{\sqrt{2}}{Z}+ \left(\frac{13}{4}+\frac{\sqrt{2}}{Z}\right) \omega^2 \tau_{nd}^2 }{\left(1+\frac{\sqrt{2}}{Z}\right)^2+ \left(\frac{449}{16}+\frac{13}{2}\frac{\sqrt{2}}{Z}+\frac{2}{Z^2}\right)\omega^2\tau_{nd}^2
+ \frac{25}{4}\omega^4\tau_{nd}^4},
\end{equation}

 \begin{equation}
 \label{b12t}
b_{1}^1 = -\frac{5}{2}\frac{\omega\tau_{nd}^2}{B}\frac{\frac{19}{4} + \frac{5}{2} \omega^2 \tau_{nd}^2 }{\left(1+\frac{\sqrt{2}}{Z}\right)^2+ \left(\frac{449}{16}+\frac{13}{2}\frac{\sqrt{2}}{Z}+\frac{2}{Z^2}\right)\omega^2\tau_{nd}^2
+ \frac{25}{4}\omega^4\tau_{nd}^4},
\end{equation}
\end{widetext}
In 3-polynomial approximation the  solution of the system (\ref{system3n})  is written as

\begin{widetext}
\begin{equation}
 \label{a03t1}
a_{0} =
\frac{165}{32}\tau_{nd}\frac{1+\frac{15\sqrt{2}}{11\,Z}-
\frac{35}{11}i\omega\tau_{nd}}
{1+\frac{61\sqrt{2}}{16\,Z}+\frac{9}{2\,Z^2}
-\left(\frac{5385}{128}+\frac{365\sqrt{2}}{32\,Z}\right)\omega^2\tau_{nd}^2-i\omega\tau_{nd}
\left(\frac{1017}{64}+\frac{667\sqrt{2}}{32\,Z}+\frac{9}{2Z^2}-\frac{175}{16}\omega^2\tau_{nd}^2\right)},
\end{equation}

\begin{equation}
 \label{a13t1}
a_{1} =
-\frac{65}{8}\tau_{nd}\frac{1+\frac{45\sqrt{2}}{52\,Z}-\frac{35}{26}\omega^2\tau_{nd}^2
-\left(\frac{713}{208}+\frac{45\sqrt{2}}{52Z}\right)i\omega\tau_{nd}}
{1+\frac{61\sqrt{2}}{16\,Z}+\frac{9}{2\,Z^2}
-\left(\frac{5385}{128}+\frac{365\sqrt{2}}{32\,Z}\right)\omega^2\tau_{nd}^2-i\omega^2\tau_{nd}^2
\left(\frac{1017}{64}+\frac{667\sqrt{2}}{32\,Z}+\frac{9}{2Z^2}-\frac{175}{16}\omega^2\tau_{nd}^2\right)},
\end{equation}
\end{widetext}

\begin{widetext}
\begin{scriptsize}
\begin{equation}
 \label{a03t}
a_{0}^1 =
\frac{165}{32}\tau_{nd}\frac{\left(1+\frac{15\sqrt{2}}{11\,Z}\right)
\left[1+\frac{61\sqrt{2}}{16\,Z}+\frac{9}{2\,Z^2}
-\left(\frac{5385}{128}+\frac{365\sqrt{2}}{32\,Z}\right)\omega^2\tau_{nd}^2\right]
+\frac{35}{11}\omega^2\tau_{nd}^2
\left(\frac{1017}{64}+\frac{667\sqrt{2}}{32\,Z}+\frac{9}{2Z^2}-\frac{175}{16}\omega^2\tau_{nd}^2\right)}
{\left[1+\frac{61\sqrt{2}}{16\,Z}+\frac{9}{2\,Z^2}
-\left(\frac{5385}{128}+\frac{365\sqrt{2}}{32\,Z}\right)\omega^2\tau_{nd}^2\right]^2+\omega^2\tau_{nd}^2
\left(\frac{1017}{64}+\frac{667\sqrt{2}}{32\,Z}+\frac{9}{2Z^2}-\frac{175}{16}\omega^2\tau_{nd}^2\right)^2},
\end{equation}

 \begin{equation}
 \label{b03t}
b_{0}^1=\frac{165}{32}\frac{\omega\tau_{nd}^2}{B} \frac{
-\frac{35}{11} \left[1+\frac{61\sqrt{2}}{16\,Z}+\frac{9}{2\,Z^2}
-\left(\frac{5385}{128}+\frac{365\sqrt{2}}{32\,Z}\right)\omega^2\tau_{nd}^2\right]
+\left(1+\frac{15\sqrt{2}}{11\,Z}\right)\left(\frac{1017}{64}+\frac{667\sqrt{2}}{32\,Z}
+\frac{9}{2Z^2}-\frac{175}{16}\omega^2\tau_{nd}^2\right)}
{\left[1+\frac{61\sqrt{2}}{16\,Z}+\frac{9}{2\,Z^2}
-\left(\frac{5385}{128}+\frac{365\sqrt{2}}{32\,Z}\right)\omega^2\tau_{nd}^2\right]^2+\omega^2\tau_{nd}^2
\left(\frac{1017}{64}+\frac{667\sqrt{2}}{32\,Z}+\frac{9}{2Z^2}-\frac{175}{16}\omega^2\tau_{nd}^2\right)^2},
\end{equation}

\begin{equation}
 \label{a13t}
a_{1}^1 =
-\frac{65}{8}\tau_{nd}\frac{\left(1+\frac{45\sqrt{2}}{52\,Z}-\frac{35}{26}\omega^2\tau_{nd}^2\right)
\left[1+\frac{61\sqrt{2}}{16\,Z}+\frac{9}{2\,Z^2}
-\left(\frac{5385}{128}+\frac{365\sqrt{2}}{32\,Z}\right)\omega^2\tau_{nd}^2\right]
+\left(\frac{713}{208}+\frac{45\sqrt{2}}{52Z}\right)\omega^2\tau_{nd}^2
\left(\frac{1017}{64}+\frac{667\sqrt{2}}{32\,Z}+\frac{9}{2Z^2}-\frac{175}{16}\omega^2\tau_{nd}^2\right)}
{\left[1+\frac{61\sqrt{2}}{16\,Z}+\frac{9}{2\,Z^2}
-\left(\frac{5385}{128}+\frac{365\sqrt{2}}{32\,Z}\right)\omega^2\tau_{nd}^2\right]^2+\omega^2\tau_{nd}^2
\left(\frac{1017}{64}+\frac{667\sqrt{2}}{32\,Z}+\frac{9}{2Z^2}-\frac{175}{16}\omega^2\tau_{nd}^2\right)^2},
\end{equation}

 \begin{equation}
 \label{b13t}
b_{1}^1=-\frac{65}{8}\frac{\omega\tau_{nd}^2}{B}
\frac{\left(1+\frac{45\sqrt{2}}{52\,Z}-\frac{35}{26}\omega^2\tau_{nd}^2\right)
\left(\frac{1017}{64}+\frac{667\sqrt{2}}{32\,Z}+\frac{9}{2Z^2}-\frac{175}{16}\omega^2\tau_{nd}^2\right)
-\left(\frac{713}{208}+\frac{45\sqrt{2}}{52Z}\right) \left[1+\frac{61\sqrt{2}}{16\,Z}+\frac{9}{2\,Z^2}
-\left(\frac{5385}{128}+\frac{365\sqrt{2}}{32\,Z}\right)\omega^2\tau_{nd}^2\right]}
{\left[1+\frac{61\sqrt{2}}{16\,Z}+\frac{9}{2\,Z^2}
-\left(\frac{5385}{128}+\frac{365\sqrt{2}}{32\,Z}\right)\omega^2\tau_{nd}^2\right]^2+\omega^2\tau_{nd}^2
\left(\frac{1017}{64}+\frac{667\sqrt{2}}{32\,Z}+\frac{9}{2Z^2}-\frac{175}{16}\omega^2\tau_{nd}^2\right)^2}.
\end{equation}
\end{scriptsize}
\end{widetext}
The values $c_0^1$ and $c_1^1$ in 2 and 3 polynomial approximations are defined using (\ref{exp_for_tens}).

The heat flux $q_i$ from (\ref{q}),(\ref{tensor nondeg}) may be written in the form

\begin{widetext}
\begin{eqnarray}
\label{qunondeg}
 q_i = -\frac{5}{2}\frac{k^{2}Tn_e}{m_e}\left[(a_{0}^{1} -  a_{1}^{1})\delta_{ik} - \varepsilon_{ikn}B_{n}(b_{0}^{1} - b_{1}^{1})   + B_{i}B_{k}(c_{0}^{1} -
 c_{1}^{1})\right]\frac{\partial T}{\partial r_k}=q_i^{(1)}+q_I^{(2)}+q_i^{(3)},
 \end{eqnarray}

\begin{eqnarray}
\label{qunondeg1}
q_i^{(1)} = -\frac{5}{2}\frac{k^{2}T n_e}{m_e}(a_{0}^{1} -  a_{1}^{1})\frac{\partial T}{\partial r_i}
=-\lambda_{nd}^{(1)}\frac{\partial T}{\partial r_i},\\
\label{qunondeg1a}
q_i^{(2)} = \frac{5}{2}\frac{k^{2}Tn_e}{m_e}\varepsilon_{ikn}B_{n}(b_{0}^{1} - b_{1}^{1})\frac{\partial T}{\partial r_k}=-\varepsilon_{ikn}B_{n}\lambda_{nd}^{(2)}\frac{\partial T}{\partial r_k},\\
\label{qunondeg1b}
q_i^{(3)} = -\frac{5}{2}\frac{k^{2}Tn_e}{m_e} B_{i}B_{k}(c_{0}^{1} -
 c_{1}^{1})\frac{\partial T}{\partial r_k}=- B_{i}B_{k}\lambda_{nd}^{(3)}\frac{\partial T}{\partial r_k}.
 \end{eqnarray}
 \end{widetext}
For 2-polynomial approximation we obtain
\newline

\begin{widetext}
\begin{eqnarray}
\label{qunondeg2}
\lambda_{nd}^{(12)}=\frac{5}{2}\frac{k^{2}T n_e}{m_e}(a_{0}^{1} -  a_{1}^{1})=\frac{25}{4}\frac{k^{2}T n_e}{m_e}\tau_{nd}\frac{\frac{5}{2}\left(1+ \frac{\sqrt{2}}{Z}\right)+ \left(-\frac{1}{2}+\frac{\sqrt{2}}{Z}\right) \omega^2 \tau_{nd}^2 }{\left(1+\frac{\sqrt{2}}{Z}\right)^2+ \left(\frac{449}{16}+\frac{13}{2}\frac{\sqrt{2}}{Z}+\frac{2}{Z^2}\right)\omega^2\tau_{nd}^2
+ \frac{25}{4}\omega^4\tau_{nd}^4},\\
\lambda_{nd}^{(22)}=-\frac{5}{2}\frac{k^{2}T n_e}{m_e}(b_{0}^{1} -  b_{1}^{1})=-\frac{25}{4}\frac{k^{2}T n_e}{m_e}\frac{\omega\tau_{nd}^2}{B}\frac{\frac{107}{8}+ \frac{3\sqrt{2}}{2Z}+ \frac{5}{2}\omega^2 \tau_{nd}^2 }{\left(1+\frac{\sqrt{2}}{Z}\right)^2+ \left(\frac{449}{16}+\frac{13}{2}\frac{\sqrt{2}}{Z}+\frac{2}{Z^2}\right)\omega^2\tau_{nd}^2
+ \frac{25}{4}\omega^4\tau_{nd}^4},\\
B^2\lambda_{nd}^{(32)}=\lambda_{nd}^{(12)}(B=0)-\lambda_{nd}^{(12)}.\qquad\qquad\qquad\qquad
\end{eqnarray}
\end{widetext}
The expressions for heat conductivity coefficients in 3-polynomial approximation are very cumbersome, and are not presented here.  They could be written explicitly  using (\ref{a03t}) - (\ref{qunondeg1b}).
\newline

Using (\ref{tensor nondeg}) we present another form of the components of the heat
conductivity tensor in the magnetic field. Three components of the heat
flux: parallel $q_{||}$,
perpendicular $q_{\perp}$ to the magnetic field $\vec B$, and
"Hall" component of the heat flux $q_{\rm hall}$,
perpendicular to both vectors $\nabla T$ and $\vec B$, with account
of (\ref{qu2t1}) or (\ref{qu3t1})  are defined by relations

$$
q_{||}=-\lambda_{||}\nabla T_{||}, $$
\begin{equation}
\label{qb2l}
\lambda_{||}
=\frac{5}{2}\frac{k^2 T n_e}{m_e}[a_0^1-a_1^1+B^2(c_0^1-c_1^1)]=\lambda_{nd},
 \end{equation}
\begin{equation}
\label{qbp2l}
 q_{\perp}=-\lambda_{\perp}\nabla T_{\perp},
\quad \lambda_{\perp} =\frac{5}{2}\frac{k^2 T
n_e}{m_e}(a_0^1-a_1^1),
\end{equation}
\begin{equation}
\label{qbh2l} q_{\rm hall}=-\lambda_{\rm
hall}\frac{\nabla T\times \vec B}{B}, \quad \lambda_{\rm
hall} =\frac{5}{2}\frac{k^2 T n_e}{m_e}B(b_0^1-b_1^1).
 \end{equation}
 The 2-polynomial results coincide with corresponding derivations obtained
 in \cite{marsh90,bk64}.

 The difference between 2 and 3 polynomial approximation may be characterised by comparison of values $Q_\perp^{(2)}$
and  $Q_\perp^{(3)}$ in Fig.1.

\begin{equation}
 \label{qperp}
Q_\perp^{(2)}=\frac{\lambda_{nd}^{(12)}}{\lambda_{nd}^{(3)}},\quad Q_\perp^{(3)}=\frac{\lambda_{nd}^{(13)}}{\lambda_{nd}^{(3)}},
\end{equation}
where $\lambda_{nd}^{(12)}$ is defined in (\ref{qunondeg2}), $\lambda_{nd}^{(3)}$ is defined in (\ref{qu3t1}), and $\lambda_{nd}^{(13)}$
is defined from (\ref{qunondeg1}),(\ref{a03t}),(\ref{a13t}) in the same way as $\lambda_{nd}^{(12)}$. The functions $Q_\perp^{(2)}(\omega\tau_{nd})$,
$Q_\perp^{(3)}(\omega\tau_{nd})$ are presented in Fig.1 for carbon, at Z=6. In this figure we have $Q_\perp^{(2)}=0.023$ and $Q_\perp^{(3)}=0.014$, at
$\omega\tau=1$.

\begin{figure}
\centerline{\hbox{\includegraphics[width=0.33\textwidth, angle=270]{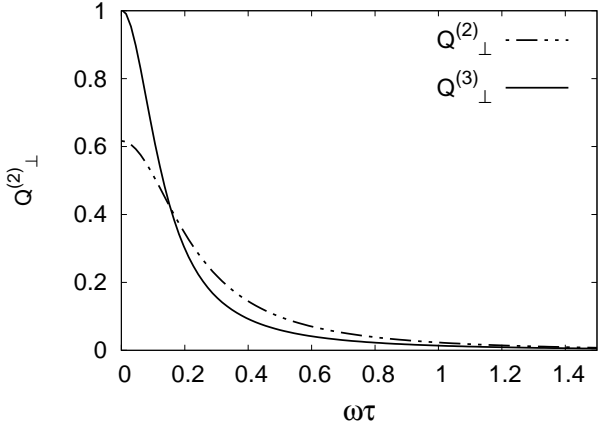}}}
\bigskip
\medskip
\caption{Comparison of 2 and 3 polynomial approximation for
nondegenerate carbon plasma at Z=6, at different $\omega\tau$.} \label{fig1}
\end{figure}

\section{Comparison of the exact solution in Lorentz approximation for a heat conductivity
with polynomial expansion}

\subsection{Exact solution in Lorentz approximation}

The Lorentz approximation for solving a kinetic equation is applied when the mass of light particles
(electrons) is much smaller than the the mass of heavy particles (nuclei), and in addition
electron-electron collisions are neglected. In this approximation the linearized Boltzmann equation has an
exact solution.
This approximation works good for metal transport coefficients,
where a strong electron degeneracy permits to neglect electron-electron collisions.
Lorenz approximation is useful for checking approximate polynomial solution, because
it gives a possibility to follow a convergence of the approximate solution to the exact one,
increasing the number of polynomials. In different approaches the solution in Lorentz
approximation was considered in \cite{chap90},\cite{schatz},\cite{wyller},\cite{wyller2}, , see also \cite{bkphys}.

The explicit exact solution in Lorentz approximation is obtained for the case $B=0$.
If we consider only the heat flux connected with the temperature gradient, at zero value of the
diffusive vector $d_i$ from (\ref{chi_a}),(\ref{chi_b}), than we obtain the expression for the
heat flux from \cite{bkphys} as

\begin{equation}
\label{qlor}
    q_i=-{640k\over \Lambda}{m_e (kT)^4\over n_NZ^2e^4h^3}
    \left(G_5-{1\over 2}\frac{G_{5/2}}{G_{3/2}} G_4\right)\frac{\partial{T}}{\partial r_{\rm i}}.
  \end{equation}
 In the limiting cases the coefficient in (\ref{qlor}) is reduces to

$$
  \lambda_{e}^l=  {40\sqrt2\over \pi^{3/2} \Lambda}k{n_e\over n_N}
    {(kT)^{5/2}\over {\rm e}^4 Z^2\sqrt{m_e}}
    = \frac{320}{3\pi}\frac{k^2 Tn_e}{m_e}\tau_{nd} \quad (ND)
$$
\begin{equation}
\label{tlor}
    \quad     ={5\over 64\Lambda}{k^2Tn_{\rm e}^2h^3\over m_{\rm e}^2n_NZ^2e^4}
    =\frac{5\pi^2}{6}\frac{k^2 Tn_e}{m_e}\tau_d \quad (D).
\end{equation}
Note, that the heat conductivity coefficient in (\ref{tlor}) determines the heat flux at zero value of the
diffusion vector $d_i=0$. Often the heat conductivity coefficient is written for the case of zero value
of the diffusion velocity $\langle v_{ \alpha i} \rangle=0$ \cite{chap90,bkphys}. When the thermal conductivity
and diffusion are calculated in the same procedure, both heat and diffusion fluxes are calculated without
any restrictions on the diffusion vector or diffusion velocity. Such consideration will be performed elsewhere.
 The
exact formulae in the Lorentz model are used \cite{chap90} for estimation of the precision of the polynomial
approximation.

The input of electron-electron collisions into the heat conductivity coefficients for different $Z$ may be estimated from the plot of normalized 3-polynomial
heat conductivity coefficients in the direction perpendicular to the magnetic field,  introducing
the value $Q_\perp^{(3l)}$, defined as

\begin{equation}
 \label{qperp1}
 Q_\perp^{(3l)}=\frac{\lambda_{nd}^{(13)}}{\lambda_{e,nd}^{l}}.
\end{equation}
Here $\lambda_{e,nd}^{l}$ is taken from the upper line in (\ref{tlor}). The curves of this value, for different Z, including Z=$\infty$, related to
Lorentz approximation, are plotted in Fig.2. The intersection of the plots with the y-axis in Fig.2 occurs in the points given in the Table 1, multiplied by
$\frac{\lambda^{(3)}_{nd}}{\lambda_{e,nd}^{l}}=0,978$. At $\omega\tau=1$ we have $Q_\perp^{(3l)}=0.0053,\,\,0.0060,\,\, 0.0083,\,\, 0.014$ for $Z=\infty,\, 26,\,6,\,2$
respectively.

\begin{figure}
\centerline{\hbox{\includegraphics[width=0.33\textwidth, angle=270]{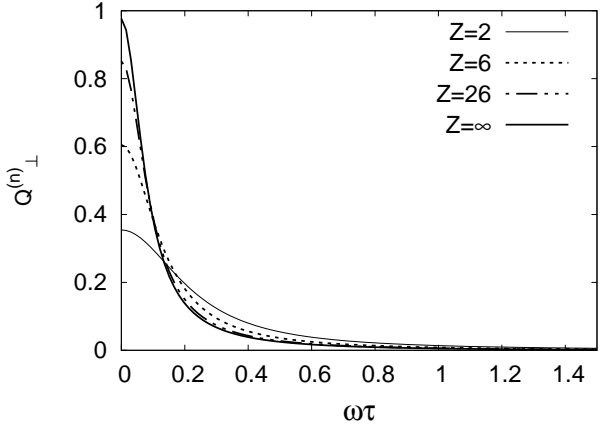}}}
\bigskip
\medskip
\caption{The plots of the value $Q_\perp^{(3l)}$ as a function of $\omega\tau$ in 3 polynomial approximation are presented
for nondegenerate  plasma of helium (Z=2), carbon (Z=6), iron (Z=26), for comparison with the
Lorentz plasma, formally corresponding to $Z=\infty$. The deviations from the Lorentz plasma are connected with the input of electron-electron collisions.
The intersection of Lorentz 3-polynomial curve (Z=$\infty$) with y-axis at 0.978 is connected with deviation from the exact solution in
Lorentz approximation.} \label{fig2}
\end{figure}

\subsection{Polynomial calculations without account of collisions between electrons}

In order to test the precision of polynomial approximation for the heat conductivity coefficients
we compare them with ones, obtained by exact solution in Lorentz approximation.
Omitting electron-electron collisions we obtain in 3 polynomial approximation the following
system

\begin{widetext}
 \begin{equation}
  \label{gsyslor}
\left\lbrace
\begin{aligned}
0  =-\frac{3}{2} i \omega n_{e}a_{0}+  a_{0}b_{00} +a_{1}b_{01}+a_{2}b_{02}\\
-\frac{15}{4} n_{e}\left( \frac{7G_{7/2}}{2G_{3/2}} -
\frac{5G_{5/2}^{2}}{2G_{3/2}^{2}}\right)   =- \frac{15}{4}  \left(
\frac{7G_{7/2}}{2G_{3/2}} - \frac{5G_{5/2}^{2}}{2G_{3/2}^{2}}\right)
i \omega
n_{e}a_{1} + a_{0}b_{10}+ a_{1} b_{11}+a_{2}b_{12}\\
0 = -
\frac{105}{16}\left(-\frac{35}{8}\frac{G_{7/2}^2}{G_{3/2}^2}+\frac{49}{2}\frac{G_{7/2}^2}{G_{5/2}^2}\frac{G_{7/2}}{G_{3/2}}
-\frac{63}{2}\frac{G_{9/2}G_{7/2}}{G_{5/2}G_{3/2}}+\frac{99}{8}\frac{G_{11/2}}{G_{3/2}}\right)
i \omega n_{e}a_{2}+a_{0}b_{20}+a_{1}b_{21}+a_{2}b_{22}
\end{aligned} \right .
 \end{equation}
 \end{widetext}

\subsection{Results for non-degenerate electrons}

In absence of the magnetic field, in Lorenz approximation with $a_{jk}=0$, the system
(\ref{general system}) is reduced to

\begin{equation}
\label{sysndb}
\left\lbrace
 \begin{aligned}
0  = a_{0}b_{00} +a_{1}b_{01}+a_{2}b_{02}\\
-\frac{15}{4} n_{e} =a_{0}b_{10}+ a_{1} b_{11}+a_{2}b_{12}\\
0 = a_{0}b_{20}+a_{1}b_{21}+a_{2}b_{22}
\end{aligned} \right .
 \end{equation}
 With account of  (\ref{b00nd})-(\ref{b22nd})this system is written as

\begin{equation}
\label{sysndb1}
\left\lbrace
 \begin{aligned}
0  = \frac{3}{2}a_{0} +\frac{9}{4}a_{1}+\frac{45}{16}a_{2}\\
-\frac{15}{4} \tau_{nd} =\frac{9}{4}a_{0}+ \frac{39}{8}a_{1}+\frac{207}{32}a_{2}\\
0 = \frac{45}{16}a_{0}+\frac{207}{32}a_{1}+\frac{1299}{128}a_{2}
\end{aligned} \right .
 \end{equation}
 This system is written for 3-polynomial approximation to the solution. Two first equations at $a_2=0$
determine the 2-polynomial approximation, giving with account of (\ref{tensor nondeg}) the following results

\begin{equation}
\label{qu2l}
a_0=\frac{15}{4}\tau_{nd},\quad a_1=-\frac{5}{2}\tau_{nd},\quad
 \end{equation}
$$\lambda^{(2)}_{ndl}=\frac{25}{4}\frac{5}{2}\frac{k^2 T n_e}{m_e}\tau_{nd}=15.63\frac{k^2 T n_e}{m_e}\tau_{nd}.
$$
In 3-polynomial approximation we obtain  the solution of
(\ref{sysndb1}) for $a_{0},\,a_{1}$, and heat conductivity
coefficient,
 with account of (\ref{tensor nondeg}), as

\begin{equation}
\label{lam3l}
a_0=\frac{165}{32}\tau_{nd},\quad
a_1=-\frac{65}{8}\tau_{nd},\quad
 \end{equation}
$$\lambda^{(3)}_{ndl}=\frac{425}{32}\frac{5}{2}\frac{k^2 T n_e}{m_e}\tau_{nd}
$$
$$
=\frac{2125}{64}\frac{k^2 T n_e}{m_e}\tau_{nd}=33.20\frac{k^2 T n_e}{m_e}\tau_{nd}.
$$
The heat coefficients obtained by the method of successive polynomial approximations should be compared with the exact solution $\lambda_{nd}^{l}$,
obtained by Lorentz method (\ref{tlor}) for non-degenerate electrons

\begin{equation}
\label{laml}
\lambda^{(l)}_{nd}=\frac{320}{3\pi}\frac{k^2 Tn_e}{m_e}\tau_{nd}=33.95\,\frac{k^2 Tn_e}{m_e}\tau_{nd}.
 \end{equation}
It is clear that the 2- polynomial solution underestimate the coefficient of the heat conductivity for more than 50\%, and
 the 3-polynomial solution differs from the exact solution only for about 2.2\%.
Equations or 3 polynomial approximation in presence of a magnetic field are obtained from (\ref{gsyslor})
 with account of (\ref{b00nd})-(\ref{b22nd}),(\ref{sysndb1})
in the form

\begin{equation}
 \label{system3n}
\left\lbrace
\begin{aligned}
0  = -\frac{3}{2} i\omega\tau_{nd} a_{0} +\frac{3}{2}a_{0} +\frac{9}{4}a_{1}+\frac{45}{16}a_{2}\\
-\frac{15}{4} \tau_{nd} =-\frac{15}{4}i\omega\tau_{nd} a_{1} +\frac{9}{4}a_{0}+ \frac{39}{8}a_{1}+\frac{207}{32}a_{2}\\
0 = -\frac{105}{16} i \omega\tau_{nd} a_{2} + \frac{45}{16}a_{0}+\frac{207}{32}a_{1}+\frac{1299}{128}a_{2}
\end{aligned} \right .
 \end{equation}
Explicit solution of equations  (\ref{system3n}) for 2 and 3 polynomial approximations is determined by formulae (\ref{a02t1})-(\ref{b13t})
at formally infinite value of $Z$.

\subsection{Partially degenerate electrons}

For partially degenerate electrons at $x_0=0$, with the degeneracy
level $DL=\frac{\varepsilon_{fe}}{kT}=1.011$, the system (\ref{gsyslor})
 is written in the form

\begin{widetext}
 \begin{equation}
  \label{gsyslor0}
\left\lbrace
\begin{aligned}
0  =-1.5 i \omega n_{e}a_{0}+  a_{0}b_{00} +a_{1}b_{01}+a_{2}b_{02}\\
-3.88 n_{e} =- 3.88 i \omega n_{e}a_{1} + a_{0}b_{10}+ a_{1} b_{11}+a_{2}b_{12}\\
0 = -7.138 i \omega n_{e}a_{2}+a_{0}b_{20}+a_{1}b_{21}+a_{2}b_{22}
\end{aligned} \right .,
 \end{equation}
 \end{widetext}
where matrix elements $b_{jk}$ are defined in (\ref{b00d0})-(\ref{b22d0}).
 In absence of a magnetic field this system is reduced to

\begin{equation}
\label{syslor0B}
\left\lbrace
 \begin{aligned}
0  = a_{0}b_{00} +a_{1}b_{01}+a_{2}b_{02}\\
-3.88 n_{e} =a_{0}b_{10}+ a_{1} b_{11}+a_{2}b_{12}\\
0 = a_{0}b_{20}+a_{1}b_{21}+a_{2}b_{22}
\end{aligned} \right .
 \end{equation}
 With account of (\ref{b00d0})-(\ref{b22d0}) this system may be written in the form

\begin{equation}
\label{sys0bl}
\left\lbrace
 \begin{aligned}
0  = 1.5 a_{0} +2.16 a_{1}+2.588 a_{2}\\
-3.88 \tau_{d0} =2.16 a_{0}+ 5.162 a_{1} +6.671 a_{2}\\
0 = 2.588 a_{0}+6.671 a_{1}+11.038 a_{2}
\end{aligned} \right .
 \end{equation}
 This system is written for 3-polynomial approximation to the solution. Two first equations at $a_2=0$
determine the 2-polynomial approximation, which with account of (\ref{tensor general}), give the following
result

\begin{equation}
\label{qu20l}
a_0=2.723\tau_{d0},\quad a_1=-1.891\tau_{d0},\quad
 \end{equation}
$$\lambda^{(2)}_{d0l}=5.043\frac{5}{2}\frac{k^2 T n_e}{m_e}\tau_{nd}=12.61\frac{k^2 T n_e}{m_e}\tau_{d0}.
$$

  In 3-polynomial approximation we obtain  the solution of
(\ref{sys0bl}) for $a_{0},\,a_{1}$, and heat conductivity
coefficient,
 with account of (\ref{tensor general}), as

\begin{equation}
\label{qu302}
  a_0=3.533\tau_{d0},\quad
 a_1=-5.295\tau_{d0},\quad
 \end{equation}

$$\lambda^{(3)}_{d0l}=8.278\frac{5}{2}\frac{k^2 T n_e}{m_e}\tau_{d0}
=22.07\frac{k^2 T n_e}{m_e}\tau_{d0}.
$$
The heat coefficients obtained by the method of successive polynomial approximations should be compared with an exact solution $\lambda_{nd}^{l}$
obtained by Lorentz method (\ref{qlor}) for non-degenerate electrons

\begin{equation}
\label{qu0l}
\lambda^{(l)}_{d0}=0.744\frac{320}{3\pi}\frac{k^2 Tn_e}{m_e}\tau_{d0}=25.26\,\frac{k^2 Tn_e}{m_e}\tau_{d0}.
 \end{equation}
It is clear that the 2- polynomial solution underestimate the coefficient of the heat conductivity for more than 50\%, and
the 3-polynomial solution differs from the exact solution for about 13\%.
 So, a convergence of the polynomial approximation
 to the exact value takes place slower than for non-degenerate electrons in previous subsection.

\bigskip

\subsection{Results for strongly degenerate electrons}

 The non-diagonal matrix
elements $b_{ik}$, $i\neq k$ for strongly
degenerate case are much smaller than the diagonal one
$b_{ii}$, according to (\ref{b00ds})- (\ref{b22ds}. In this case at $x_0\gg 1$, and
neglecting terms $\sim 1/x_0$  we obtain
 a simplified system (\ref{gsyslor}) for 3-polynomial expansion as

\begin{widetext}
 \begin{equation}
 \label{system3d}
\left\lbrace
\begin{aligned}
0  =-\frac{3}{2} i \omega n_{e}a_{0}+  a_{0}b_{00}\qquad\qquad \\
-\frac{15}{4} n_{e}\frac{2\pi^2}{15}
 =- \frac{15}{4} \frac{2\pi^2}{15}i \omega
n_{e}a_{1}+ a_{1} b_{11} \qquad\qquad\\
0 =  -
\frac{105}{16}\left(-\frac{35}{8}\frac{G_{7/2}^2}{G_{3/2}^2}
+\frac{49}{2}\frac{G_{7/2}^2}{G_{5/2}^2}\frac{G_{7/2}}{G_{3/2}}
-\frac{63}{2}\frac{G_{9/2}G_{7/2}}{g_{5/2}G_{3/2}}+\frac{99}{8}\frac{G_{11/2}}{G_{3/2}}\right)
i \omega n_{e}a_{2}+a_{2}b_{22}
\end{aligned} \right .
 \end{equation}
 \end{widetext}
Solution of the system (\ref{system3d}) is
written in the form, with account of (\ref{gammaexp}),(\ref{b01ds})

\begin{equation}
\label{sd1}
a_0=0,\quad a_2=0,
\end{equation}
\begin{widetext}
\begin{equation}
\label{sd2}
 a_1=\frac{\frac{15}{4} n_{e} \frac{2\pi^2}{15}
 }{\frac{15}{4}\frac{2\pi^2}{15} i n_{e}\omega- b_{11}}
=\frac{\frac{\pi^2 n_e}{2}}
{i\omega \frac{\pi^2 n_e}{2}- \frac{\pi^2 n_e}{2\tau_d}}=-\frac{\tau_d}{1-i\omega\tau_d}=
-\tau_d\frac{1+i\omega\tau_d}{1+\omega^2\tau_d^2}=a_1^{1}+iB b_1^{1},
 \end{equation}
 \end{widetext}

 \begin{equation}
\label{sd3}
 c_1^1=-\frac{\tau_d}{B^2} \frac{\omega^2\tau_d^2}{1+\omega^2\tau_d^2}.
 \end{equation}

Using (\ref{tensor general}) we obtain the components of the heat
conductivity tensor in the magnetic field for strongly degenerate electrons
in polynomial approximation. 3 components of the heat
flux: parallel $q_{||}^{sd}$,
perpendicular $q_{\perp}^{(sd)}$ to the magnetic field $\vec B$, and
"Hall" component of the heat flux $q_{\rm hall}^{(sd)}$,
perpendicular to both vectors $\nabla T$ and $\vec B$, with account
of (\ref{qlor}),(\ref{sd1})-(\ref{sd2})  are defined by relations

$$
q_{||}^{(sd)}=-\lambda_{||}^{(sd)}\nabla T_{||}, $$
\begin{equation}
\label{sd4}
\lambda_{||}^{(sd)}
=-\frac{\pi^2}{3}\frac{k^2 T n_e}{m_e}(a_1^1+B^2\,c_1^1)=\frac{\pi^2}{3}\frac{k^2 T n_e}{m_e}\tau_d.
 \end{equation}
\begin{equation}
\label{sd5}
 q_{\perp}^{(sd)}=-\lambda_{\perp}^{(2)}\nabla T_{\perp},
\quad \lambda_{\perp}^{(sd)} =-\frac{\pi^2}{3}\frac{k^2 T n_e}{m_e}a_1^1
\end{equation}
$$=\frac{\pi^2}{3}\frac{k^2 T n_e}{m_e}\frac{\tau_d}{1+\omega^2\tau_d^2}
$$
\begin{equation}
\label{sd6}
 q_{\rm hall}^{(sd)}=-\lambda_{\rm
hall}^{(2)}\frac{\nabla T\times \vec B}{B}, \quad \lambda_{\rm
hall}^{(sd)} =-\frac{\pi^2}{3}\frac{k^2 T n_e}{m_e}B\,b_1^1.
 \end{equation}
Comparing (\ref{sd6}) with the exact value for strongly degenerate electrons from Lorentz
approximation (\ref{tlor}) we see, that polynomial approximation, where the terms $\sim x_0^{-1}$
 are neglected, gives is 2.5 times smaller value than the exact one. This value, as well as a simple
 dependence of the heat conductivity tensor on the magnetic field $q_{||}^{(sd)}/q_{\perp}^{(sd)}=
 (1+\omega^2\tau_d^2)$ follows also from a rough theory of heat conductivity and diffusion in
 presence of a magnetic field, based on the mean free path, which is described in \cite{chap90}.
 The value of the heat conductivity coefficient, following from this approach, was considered
 in \cite{fl76},\cite{yau80}, and many subsequent papers.
    As mentioned above the heat flux calculated here is connected only with a temperature gradient,
when the diffusion vector $d_i=0$. In laboratory conditions when the electrical conductivity is small and
electrical current is damped rapidly, another limiting case is considered, where $j_i\sim \langle v_i\rangle=0$.
This restriction leads to linear connection between  $d_i$ and $\nabla T$, what permits \cite{chap90} to exclude $d_i$ and to express
the heat flux as directly proportional to $\nabla T$, with another heat conductivity coefficient $\lambda_j$. For
strongly degenerate electrons we have $\lambda_j=0.4\lambda_e^l=\lambda_{||}^{(sd)}$, see (\ref{tlor}),(\ref{sd4}) and \cite{bkphys}.

\bigskip

\section{Heat conductivity of strongly degenerate electrons in presence of magnetic field: Lorentz approximation}

The equation for the function $\xi$ from
(\ref{chi_a}),(\ref{eq_sys}) may be written in the form, using
relations $f_0'=f_0$,  $\xi'=\xi$, $u'_i=u_i\,cos\theta$, and making
integration over  $dc_{Ni}$ with account of (\ref{nuclear_norm}), as

\begin{widetext}
\begin{eqnarray}
\label{eqlor}
 f_{0}(1-f_{0})(u^{2} - \frac{5G_{5/2}}{2G_{3/2}})  =
-iBf_{0}(1-f_{0})\frac{e \xi}{m_{e} c}u_{i}+ f_{0}(1-f_{0})n_N\xi
 \int(1-\cos\theta) g_{eN}bdbd\varepsilon.
\end{eqnarray}
\end{widetext}
The function $\xi$ is defined by expression

\begin{equation}
\label{xilor}
 \xi=\frac{u^2-\frac{5}{2}\frac{G_{5/2}}{G_{3/2}}}
 {2\pi n_N \int_0^\infty (1-\cos\theta)g b db  -
 i\omega}.
\end{equation}
 Using
(\ref{omegap1})-(\ref{omegap5}) we obtain in Lorenz approximation,
with $g_{12}=v$,

\begin{equation}
\label{xilor1}
 \int_0^\infty (1-\cos\theta)g b db =2\frac{e^4
 Z^2}{m_e^2 v^3}\Lambda
\end{equation}

\begin{equation}
 \label{xilor2}
 \xi=\frac{u^2-\frac{5}{2}\frac{G_{5/2}}{G_{3/2}}}
 {4\pi n_N \left(\frac{m_e}{2kT}\right)^{3/2}\frac{e^4
 Z^2}{m_e^2 u^3}\Lambda - i\omega}.
\end{equation}
According to (\ref{xicomp}) we have

\begin{equation}
 \label{xilor3}
\xi=A^{(1)}+iB A^{(2)},
\end{equation}

\begin{equation}
 \label{xilor4}
 A^{(1)}=\frac{(u^2-\frac{5}{2}\frac{G_{5/2}}{G_{3/2}})
 4\pi n_N \left(\frac{m_e}{2kT}\right)^{3/2}\frac{e^4
 Z^2}{m_e^2 u^3}\Lambda}
 {\left[4\pi n_N \left(\frac{m_e}{2kT}\right)^{3/2}\frac{e^4
 Z^2}{m_e^2 u^3}\Lambda\right]^2 + \omega^2}.
\end{equation}

\begin{equation}
 \label{xilor5}
  A^{(2)}=\frac{\omega}{B}\frac{u^2-\frac{5}{2}\frac{G_{5/2}}{G_{3/2}}}
 {\left[4\pi n_N \left(\frac{m_e}{2kT}\right)^{3/2}\frac{e^4
 Z^2}{m_e^2 u^3}\Lambda\right]^2 + \omega^2}.
\end{equation}

\begin{equation}
 \label{xilor6}
 A^{(3)}=A^{(1)}(B=0)-A^{(1)}.
\end{equation}
 The expression for the heat flux,
following from (\ref{qi}),(\ref{chi}),(\ref{chi_a}),(\ref{xicomp}),
(\ref{xilor2})-(\ref{xilor6}) is  written as

\bigskip

\begin{widetext}
\begin{eqnarray}
\label{xilor7}
q_i=-\frac{2\pi}{3}\frac{m_e^4}{h^3T}\left(\frac{2kT}{m_e}\right)^{7/2}\left[\delta_{ij}\int_0^\infty
f_0(1-f_0)A^{(1)}x^{5/2}dx -\varepsilon_{ijk}B_k\int_0^\infty
f_0(1-f_0)A^{(2)}x^{5/2}dx \right.\nonumber\\
\left.+ B_i B_j \int_0^\infty
f_0(1-f_0)A^{(3)}x^{5/2}dx\right]\frac{\partial T}{\partial x_j}=
 q_i^{(1)}+q_i^{(2)}+q_i^{(3)}, \qquad x=u^2, \qquad \nonumber\\
 q^{(1)}_i=-\frac{2\pi}{3}\frac{m_e^4}{h^3T}\left(\frac{2kT}{m_e}\right)^{7/2}\int_0^\infty
f_0(1-f_0)A^{(1)}x^{5/2}dx \frac{\partial T}{\partial x_i}=-\lambda^{(1)}_{sd} \frac{\partial T}{\partial x_i},\qquad \\
 q^{(2)}_i=\varepsilon_{ijk}B_k
\frac{2\pi}{3}\frac{m_e^4}{h^3T}\left(\frac{2kT}{m_e}\right)^{7/2}\int_0^\infty
 f_0(1-f_0)A^{(2)}x^{5/2}dx\frac{\partial T}{\partial x_j}=
 -\varepsilon_{ijk}B_k \lambda^{(2)}_{sd}\frac{\partial T}{\partial x_j}, \qquad \nonumber\\
 q_i^{(3)}=-B_i B_j \frac{2\pi}{3}\frac{m_e^4}{h^3T}\left(\frac{2kT}{m_e}\right)^{7/2}
 \int_0^\infty
 f_0(1-f_0)A^{(3)}x^{5/2}dx \frac{\partial T}{\partial x_j}=
 -B_i B_j\lambda^{(3)}_{sd} \frac{\partial T}{\partial x_j},\qquad \nonumber
\end{eqnarray}
\end{widetext}
 For strongly degenerate electrons at $x_0\gg 1$ the
integrals in (\ref{xilor7}) with $A^{1},\,\, A^{2}\,\,A^{(3)}$ from
(\ref{xilor4})-(\ref{xilor6}), are expressed analytically , using
expansion formula \cite{dau90})

\begin{equation}
 \label{xilor8}
 \int_0^\infty
 \frac{f(x)dx}{e^{x-x_0}+1}=\int_0^{x_0}f(x)dx+\frac{\pi^2}{6}f^{'}(x_0)+...
\end{equation}
After partial integration we obtain the expression, which are
suitable for integration by formula (\ref{xilor8})

\begin{equation}
 \label{xilor9}
 \lambda^{(1)}=\frac{2\pi}{3}\frac{m_e^4}{h^3T}\left(\frac{2kT}{m_e}\right)^{7/2}\int_0^\infty
 f_0\frac{d\,(A^{(1)}x^{5/2})}{dx} dx
\end{equation}

\begin{equation}
 \label{xilor10}
 \lambda^{(2)}=-\frac{2\pi}{3}\frac{m_e^4}{h^3T}\left(\frac{2kT}{m_e}\right)^{7/2}\int_0^\infty
 f_0\frac{d\,(A^{(2)}x^{5/2})}{dx} dx
\end{equation}

\begin{eqnarray}
 \label{xilor11}
B^2 A^{(3)}=A^{(1)}(B=0)-A^{(1)}, \nonumber\\
B^2 \lambda^{(3)}=\lambda^{(1)}(B=0)-\lambda^{(1)}.
\end{eqnarray}
Applying (\ref{xilor8}) to the integrals
(\ref{xilor9}),(\ref{xilor10}), we obtain

\begin{widetext}
\begin{equation}
 \label{xilor12}
\lambda^{(1)}=\frac{2\pi}{3}\frac{m_e^4}{h^3
T}\left(\frac{2kT}{m_e}\right)^{7/2}\left[
A^{(1)}(x_0)x_0^{5/2}+\frac{\pi^2}{6}\frac{d^2\,(A^{(1)}x^{5/2})}{dx^2}|_{x=x_0}\right],
\end{equation}

\begin{equation}
 \label{xilor13}
\lambda^{(2)}=-\frac{2\pi}{3}\frac{m_e^4}{h^3
T}\left(\frac{2kT}{m_e}\right)^{7/2}\left[
A^{(2)}(x_0)x_0^{5/2}+\frac{\pi^2}{6}\frac{d^2\,(A^{(2)}x^{5/2})}{dx^2}|_{x=x_0}\right],
\end{equation}
Using (\ref{xilor4}),(\ref{xilor5}), and writing the formula using
$\tau_{d}$ from (\ref{taud}, we write the heat conductivity
coefficients in the form

\begin{equation}
 \label{xilor14}
\lambda^{(1)}=\frac{5\pi^2}{6}\frac{k^2 T n_e}{m_e }\tau_d
\left\{\frac{1}{1+\omega^2\tau_d^2}-
\frac{6}{5}\frac{\omega^2\tau^2_d}{(1+\omega^2\tau_d^2)^2}
 -\frac{\pi^2}{10}\left[\frac{1}{1+\omega^2\tau_d^2\left(\frac{x^3}{x_0^3}\right)}\right]^{''}
|_{x=x_0}\right\},
\end{equation}

\begin{equation}
 \label{xilor15}
\lambda^{(2)}=-\frac{4\pi^2}{3}\frac{k^2 T n_e}{m_e }\frac{\tau_d^2\omega}{B}
\left\{\frac{1}{1+\omega^2\tau_d^2}-
\frac{3}{4}\frac{\omega^2\tau^2_d}{(1+\omega^2\tau_d^2)^2}
 -\frac{\pi^2}{16}\left[\frac{1}{1+\omega^2\tau_d^2\left(\frac{x^3}{x_0^3}\right)}\right]^{''}
|_{x=x_0}\right\},
\end{equation}
\end{widetext}
In the case of strongly degenerate electrons the equations
(\ref{xilor4})-(\ref{xilor7}),(\ref{xilor14},(\ref{xilor15}) give an asymptotically
exact solution for the heat conductivity coefficients, because
collisions between electrons can be neglected in this case.
  The difference between the exact $[\lambda^{(1)}]/[\lambda^{(1)}(B=0)]$ from (\ref{xilor14}), and phenomenological (\ref{firstone})
account of the magnetic field influence on the heat conductivity
  coefficients is presented in Fig.3. Here the ratios between the values, which are perpendicular and parallel to magnetic field, are plotted for $kT=0.09E_{f}$.
  At $\omega\tau=1.5$ the exact value  of this ratio is 4 times smaller than the phenomenological one.

The heat flux defined in (\ref{xilor7})-(\ref{xilor15}) corresponds to the
situation when the diffusion vector $d_1$ from (\ref{chi_b}) is zero. In general case the heat and diffusion (electrical current) fluxes
are connected with each other by diffusion vector $d_i$ and temperature gradient $\partial T/\partial x_i$ \cite{chap90}.

\begin{figure}[h!]
\centerline{\hbox{\includegraphics[width=0.33\textwidth, angle=270]{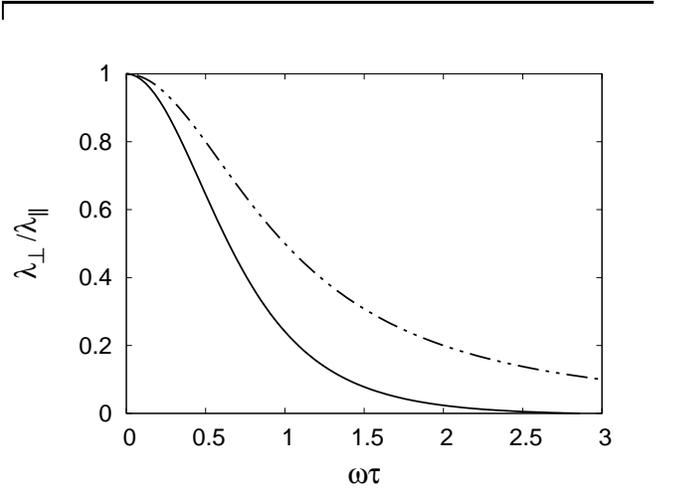}}}
\caption{The plots of the ratio $\lambda_{\perp}/\lambda_{\parallel}$ as a function of $\omega \tau$ are presented for phenomenologically obtained heat conductivity  (dash-dot line)  for comparison with heat conductivity obtained by the solution of Boltzmann equation in Lorentz appoximation (solid line) with $kT=0.09E_{f}$. \label{fig3}}
\end{figure}

\section{Discussion}

In this paper a thermal conductivity tensor is found for arbitrary degenerate non-relativistic
electrons in presence of a non-quantizing magnetic field. For nondegenerate electrons
the conductivity tensor is derived from the solution of
a Boltzmann kinetic equation by classical Chapman - Enskog method using an expansion on the
Sonyne polynomials, with remaining two and three terms. The electron-nucleus and electron-electron collisions
are taken into consideration. The tensor of the thermal conductivity is written for arbitrary local direction of the
magnetic field and the temperature gradient, in the Cartesian coordinate system, following \cite{bkl88}. Our
results exactly coincide with the results of previous authors \cite{lands51,marsh90,bk64} in two polynomial case.
The analytic solution in three polynomial approximation was not obtained before.
 The value of the thermal conductivity coefficient obtained in the well-known work
of Braginskii \cite{brag57} in two-polynomial approximation
 is two times less than our correspondong value. It is connected with an approach used in the paper \cite{brag57},
which is different from the classical Chapman - Enskog method \cite{chap90}. In his consideration one half of the
thermal flux is hidden inside the so-called "thermal force", so that the resulting heat flux in the co-moving frame
is the same in both considerations.
  The heat conductivity coefficients for strongly degenerate electrons, in presence of magnetic field, are obtained
asymptotically exactly in Lorentz approximation, when the electron-electron collision may be neglected in comparison with
electron-nuclei collisions at nondegenerate nuclei.

In most works considering the heat conductivity in astrophysical objects, in the neutron stars in particular,
following Flowers and Itoh \cite{fl76}, the
influence of the magnetic field on the heat flux was taken into account  phenomenologically using the coefficient
$1/(1+\omega^2\tau^2)$, which decreases the heat flux in the direction perpendicular to the direction of a
magnetic field. Our results, obtained by the solution of Boltzmann equation show, that the influence of the magnetic field
on the coefficients of heat conductivity is stronger, and has a  more complicated character Fig.3.

On the example of the Lorentz approximation it was shown that the precision of approximation by the raw of the orthogonal functions,
analogous to Sonyne polynomials,
decreases with increasing of the degeneracy level. For strongly degenerate electrons the number of functions, needed for
good precision is increasing $\sim x_0$, at $x_0 \gg 1$, and for small number of number of functions the resulting heat conductivity
coefficient at $B=0$ is 2.5 times smaller than the exact value. For moderately degenerate electrons with $x_0=0$ the approximation by three
orthogonal functions gives the value of the heat conductivity coefficient approximately 13\% smaller than the exact value (in Lorentz approximation, at $B=0$).
In the same approximation for nondegenerate electrons the value of the heat conductivity coefficient is only 2.2\% smaller than the exact value.
Note that the electron-electron collision even more decrease the the value of the heat conductivity coefficients. Therefore, in three functional
approximation, the Lorentz approach may give a more exact value for heat conductivity coefficient of moderately degenerate electrons, than with account of
electron-electron collisions. The simple linear interpolation, between "exact" results for nondegenerate, and strongly degenerate electrons,
may be suggested for all heat conductivity coefficients $\lambda_i(x_0)$ in presence of the magnetic field as

\begin{equation}
 \label{interpn}
 \lambda_i(x_0)=\lambda_i^{(nd)}\frac{1-x_0}{2-x_0}+\lambda_i^{(sd)}\frac{1}{2-x_0},\quad {\rm at}\quad x_0\le 0,
 \end{equation}

 \begin{equation}
 \label{interpd}
 \lambda_i(x_0)=\lambda_i^{(nd)}\frac{1}{2+x_0}+\lambda_i^{(sd)}\frac{1+x_0}{2+x_0},\quad {\rm at}\quad x_0\ge 0.
 \end{equation}
 Here the indices (nd), (sd) are related to nondegenerate and strongly degenerate values, respectively.

 The Chapmen-Enskog method could be used for sufficiently dense gas (plasma), where the time between collisions of particles is
 the smallest among other characteristic times. In presence of a magnetic field we have, in addition to the time of the  system flyover, and
 characteristic time of the parameter variations in plasma, the time  of the flight over Larmor circle $\tau_L=\frac{2\pi}{\omega}$. This time should be
 much less than $\tau$, equal to $\tau_{nd}$ or $\tau_d$, what leads to inequality, at which the Chapmen-Enskog method could be used, in the form

 \begin{equation}
 \omega\tau\ll 2\pi.
 \end{equation}
 Therefore the consideration in this paper could be safely applied at $\omega\tau \lesssim 1$, and for larger $\omega\tau$ only qualitative
 estimations could be obtained.

Our calculations have been done for non-relativistic electrons, while in deep layers of the neutron star
crust the relativistic effects become important. The main relativistic effect of increasing the effective electron mass
may be taken into account approximately, following \cite{yau80}, by writing in all expressions the relativistic electron mass
$m_{e*}=(m_e^2+p_{Fe}^2/c^2)^{1/2}$  instead of the rest mass $m_e$. The account of quantum effects is connected with
consideration of discrete Landau levels in strong magnetic fields.  This complicated problem is not yet solved.

The transport coefficients calculated here determine a heat flux carried by electrons in the case of
zero diffusion vector $d_i$. In a general case of nonzero  diffusion vector $d_i$ and temperature gradient $\partial T/\partial x_i$,
 the heat and diffusion (electrical current) fluxes
are connected with each other, and are defined by 4 kinetic coefficients \cite{chap90}, having a tensor structure in presence of a magnetic field.
The general consideration of heat and electrical conductivity of degenerate electrons will be done elsewhere.

The new coefficients can be used  for calculation of temperature distribution in white dwarfs, on the surface and in the crust of magnetized neutron star, as well as in the magnetized matter accreting to the magnetized neutron star.
The temperature distribution over the surface of NS is important for understanding of the geometry of magnetic field  inside the neutron star and near its surface.

\begin{acknowledgments}
The work of GSBK and MVG was  supported
 by the Russian Science Foundation  grant No. 15-12-30016.
\end{acknowledgments}

\end{document}